\documentclass[lettersize,journal]{IEEEtran}
\usepackage{amsmath,amsfonts}
\usepackage{algorithmic}
\usepackage{algorithm}
\usepackage{array}
\usepackage[caption=false]{subfig}
\usepackage{textcomp}
\usepackage{stfloats}
\usepackage{url}
\usepackage{verbatim}
\usepackage{graphicx}
\usepackage{cite}
\usepackage[table]{xcolor}
\usepackage{pifont}
\usepackage{enumitem}
\usepackage{marvosym}

\newcommand{\sysname}{{SEMSO}}

\begin{document}

\title{{\sysname}: A Secure and Efficient Multi-Data Source Blockchain Oracle}

\author{Youquan Xian$^*$ ~\IEEEmembership{Graduate Student Member,~IEEE}, Xueying Zeng$^*$ ~\IEEEmembership{Graduate Student Member,~IEEE}, Chunpei Li, Peng Wang, Dongcheng Li, Peng Liu\textsuperscript{\Letter}, and Xianxian Li

\thanks{Youquan Xian, Xueying Zeng, Chunpei Li, Peng Wang, Dongcheng Li, Peng Liu, Xianxian Li are affiliated with both the Key Lab of Education Blockchain and Intelligent Technology, Ministry of Education, and the Guangxi Key Lab of Multi-Source Information Mining and Security at Guangxi Normal University, Guilin, 541004, China. (e-mail: 
xianyouquan@stu.gxnu.edu.cn; 
xyz@stu.gxnu.edu.cn;
licp@gxnu.edu.cn;
wangp@gxnu.edu.cn;
ldc@gxnu.edu.cn;\textit{}
liupeng@gxnu.edu.cn; lixx@gxnu.edu.cn), 
$^*$Authors Youquan Xian and Xueying Zeng have made equal contributions to this research.
\textsuperscript{\Letter}Peng Liu is the corresponding author.}

\thanks{Manuscript created October, 2022.}}

\markboth{Journal of \LaTeX\ Class Files,~Vol.~14, No.~8, August~2021}%
{Shell \MakeLowercase{\textit{et al.}}: A Sample Article Using IEEEtran.cls for IEEE Journals}

\maketitle

\begin{abstract}
In recent years, blockchain oracle, as the key link between blockchain and real-world data interaction, has greatly expanded the application scope of blockchain. In particular, the emergence of the Multi-Data Source (MDS) oracle has greatly improved the reliability of the oracle in the case of untrustworthy data sources. However, the current MDS oracle scheme requires nodes to obtain data redundantly from multiple data sources to guarantee data reliability, which greatly increases the resource overhead and response time of the system. Therefore, in this paper, we propose a Secure and Efficient Multi-data Source Oracle framework ({\sysname}), which nodes only need to access one data source to ensure the reliability of final data. First, we design a new off-chain data aggregation protocol TBLS, to guarantee data source diversity and reliability at low cost. Second, according to the rational man assumption, the data source selection task of nodes is modeled and solved based on the Bayesian game under incomplete information to maximize the node's revenue while improving the success rate of TBLS aggregation and system response speed. Security analysis verifies the reliability of the proposed scheme, and experiments show that under the same environmental assumptions, {\sysname} takes into account data diversity while reducing the response time by 23.5\%.
\end{abstract}


\section{Introduction}
\IEEEPARstart{I}{n} recent years, with the rapid development of blockchain technology, its applications have become increasingly widespread in areas such as Decentralized Finance (DeFi) \cite{schar2021decentralized}, supply chain management \cite{zhu2024using}, and healthcare \cite{andrew2023blockchain}. However, the closed nature of blockchain networks makes it difficult to directly access and process real-world data, limiting their potential and scope in practical applications. To address this limitation, blockchain oracles have emerged as a critical link facilitating the interaction between blockchain and real-world data, significantly advancing the application and development of blockchain technology \cite{patel2021blockchain,hassan2023trust}.

In the early days, oracle schemes usually used voting games \cite{peterson2015augur, berryhill2019astraea}, threshold signature \cite{liu2024trustworthy, lin2022novel}, reputation mechanisms \cite{breidenbach2021chainlink,xian2024distributed}, etc. methods to construct a distributed oracle node trust system, or used a Trusted Execution Environment (TEE) \cite{zhang2016town, liu2022extending} and improved protocol of TLS (such as TLS-N \cite{ritzdorf2018tls}) to ensure the reliability of data acquisition. However, it is essential to acknowledge that data sources may also be untrustworthy. Blindly trusting data from a Single Data Source (SDS) can lead to significant financial losses \footnote{In 2019, Synthetix, a DeFi project on Ethereum, experienced an incident where a commercial API intermittently reported an exchange rate for the Korean won 1,000 times higher than the actual rate. This erroneous data was adopted by the price feed contract, resulting in financial implications amounting to nearly \$1 billion. \url{https://blog.synthetix.io/response-to-oracle-incident/}} \cite{iqbal2024bridging,oracle_news}.

\begin{figure}[h]
  \centering
  \includegraphics[width=0.9\linewidth]{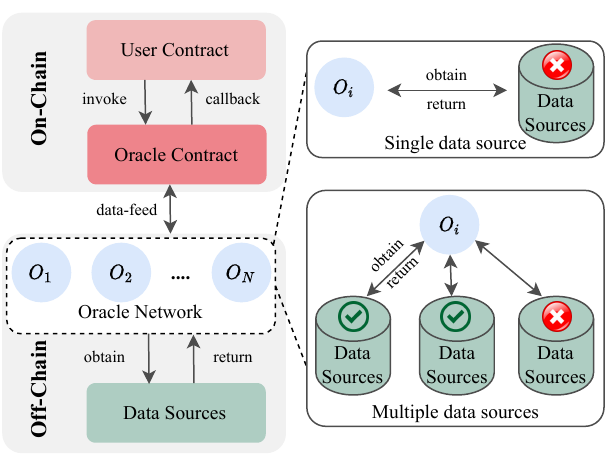}
  \caption{SDS and MDS oracle.}
  \label{fig:oracle}
\end{figure}

While Multi-Data Sources (MDS) oracle solutions enhance data reliability by establishing a distributed trust system through the diversification of data sources, they also introduce significant challenges \cite{lv2021blockchain,almi2023graph,gigli2023decentralized,dong2023daon,xiao2023decentralized}. To ensure the diversity of data sources, individual nodes are required to access MDS, which results in an increased consumption of system resources, additional costs, and response times. This issue is particularly pronounced in scenarios where the access costs of data sources are high, leading to a substantial reduction in the willingness of both nodes and users to participate. Such drawbacks not only negatively impact user experience but also impede the practical adoption of oracle technology. Fig. \ref{fig:oracle} illustrates the fundamental structures and distinctions between SDS and MDS oracles.

Although solutions such as IoT \cite{xian2024distributed} try to allocate fast response nodes to task data sources through reputation mechanisms to improve the response speed. However, it relies on the response time of TEE-trusted record nodes and uses smart contracts to calculate and store reputation values. This leads to a lot of additional overhead, and it is difficult to ensure the diversity of data sources. Therefore, the core challenge is whether it is possible to access a single data source only through nodes without relying on TEE and smart contracts while ensuring the ultimate reliability of the data.

To address the aforementioned problem, this study introduces a secure and efficient multi-source oracle framework {\sysname}. It significantly optimizes data access frequency and system response speed while maintaining the diversity of data sources. Firstly, we present a novel off-chain data aggregation protocol, TBLS, which is designed to ensure the final diversity of data sources with minimal resource expenditure. Secondly, leveraging the TBLS protocol and the assumption of rational agents, we model the nodes using a Bayesian game framework. It aims to maximize node rewards while concurrently enhancing the success rate of TBLS aggregation and improving overall system response times. Table \ref{tab:experimental-setup} highlights the distinctions between the proposed framework and previous solutions.

\begin{table}[h]
\caption{Comparison of the proposed scheme with previous approaches}
\resizebox{\linewidth}{!}{%
\begin{tabular}{ccccc}
\hline
Scheme & Untrusted Data Sources & Number of Data Accesses & Response Speed & Diversity of Data Sources \\ \hline
SDS Schemes & \ding{56} & $O(N)$ & Fast & \ding{56} \\
MDS Schemes & \ding{52} & $O(N \times M)$ & Slow & \ding{52} \\
\rowcolor{lightgray} Ours & \ding{52} & $O(N)$ & Fast & \ding{52} \\ \hline
\end{tabular}
}

\label{tab:experimental-setup}
\end{table}

The main contributions of this paper are as follows: 

\begin{itemize} 
\item We designed a data aggregation protocol, TBLS, which integrates TLS-N with threshold signatures to ensure data source diversity and reliability while maintaining low resource expenditure. 
\item Based on a Bayesian game model, we optimized the data source selection strategy for off-chain nodes, resulting in enhanced data aggregation success rates and improved system response times. 
\item Both experimental and security analyses validate the reliability of the proposed framework. Notably, the approach not only maintains data diversity but also achieves a 23.5\% reduction in response times while significantly decreasing overall system resource consumption.
\end{itemize}

The remainder of this paper is structured as follows: Section \ref{bg} introduces the relevant background knowledge and existing challenges. Section \ref{main} offers a brief description of the workflow of the proposed framework \sysname. Section \ref{detailed} presents the implementation details of the proposed solution. Section \ref{experiment} discusses the experimental results and security analysis. Finally, Section \ref{conclusion} concludes the paper and outlines future research directions.

\section{RELATED WORK AND EXISTING CHALLENGES}
\label{bg}
Blockchain oracles serve as a trusted bridge connecting blockchains with the external world, providing blockchains with access to external data that they cannot retrieve autonomously, such as weather conditions, prices, exchange rates, etc. It extends the range of blockchain applications \cite{pasdar2023connect,lo2020reliability}.

To tackle the issue of ensuring the trustworthiness of data obtained by oracles, projects like Augur \cite{peterson2015augur} and Astraea \cite{berryhill2019astraea} introduced mechanisms where multiple oracle nodes vote and engage in game theory, placing bets on the veracity of the data. Deepthought \cite{gennaro2022deepthought} extended Astraea`s voting system by linking it to user reputation, incentivizing the most honest users while reducing the risk of corruption caused by adversaries or passive voters, thereby improving data credibility. With the growing popularity of Trusted Execution Environments (TEE), researchers such as Zhang et al. \cite{zhang2016town}, Liu et al. \cite{liu2022extending}, and Woo et al. \cite{woo2020distributed} have combined TEE with oracles to ensure the integrity and consistency of the data being retrieved. However, all of these approaches inherently assume the trustworthiness of the data sources, making it challenging to ensure their effectiveness when the data sources themselves are untrustworthy.

To mitigate reliance on an SDS, MDS oracles have emerged. By integrating MDS, they enhance data reliability in scenarios where data sources may be untrustworthy. Researchers like Lv et al. \cite{lv2021blockchain} and Almi’Ani et al. \cite{almi2023graph} have employed reputation mechanisms and weighted graphs to quantify the trustworthiness of data sources, thereby recommending reliable sources for tasks. Gigli et al. \cite{gigli2023decentralized} proposed the DESMO architecture, which assigns reputation scores to data sources to select multiple trustworthy ones for data retrieval. DAON \cite{dong2023daon} utilizes distributed data sources when the trustworthiness of individual sources is indeterminable, collecting data by querying a set of sources and employing strategies such as majority voting to obtain a single answer. Xiao et al. \cite{xiao2023decentralized} addressed situations where both oracle nodes and data sources are untrustworthy by using a two-stage truth discovery process to approximate the true data values. While MDS oracles increase data credibility by diversifying data sources and negating the impact of a few malicious ones, they also require oracle nodes to redundantly fetch data from multiple data sources, significantly increasing resource consumption and response times, which can decrease participation enthusiasm and system efficiency.

Although reputation mechanisms \cite{breidenbach2021chainlink,xian2024distributed} and threshold signatures \cite{liu2024trustworthy,lin2022novel} are frequently employed to incentivize rapid responses from oracle nodes, these methods do not ensure the diversity and authenticity of data sources. While oracles can utilize improved TLS-based protocols, such as TLS-N, to obtain data from sources and provide third-party verification of the data`s authentic origin \cite{zhang2020deco, ritzdorf2018tls,luo2024proxying}, their effectiveness remains fundamentally dependent on the trustworthiness of the data sources.

Many scholars have conducted in-depth research on blockchain oracles. Nevertheless, there remain two critical challenges that urgently need to be addressed:
\paragraph{Challenge 1} It is imperative that Oracle develops a novel off-chain data aggregation protocol that can not only guarantee the dependability of data in the form of a threshold signature but also authenticate the data's provenance and guarantee the diversity and reliability of the data.
\paragraph{Challenge 2} The method of node redundancy traversing all data sources is inefficient, and random access to a single data source cannot ensure the diversity of the final data. Therefore, a reasonable data source selection strategy is required to guarantee the diversity and response speed of the final data source when the node only accesses one data source.

\section{{\sysname} OVERVIEW}
\label{main}
The process of an oracle task typically begins with a user contract (e.g., a currency exchange contract). The user contract calls the oracle contract interface to trigger a data request event (e.g., the current exchange rate) and pays a service fee. The oracle contract receives the request task and writes it into the blockchain event. Off-chain oracle nodes $\{\mathcal{O}_1,...,\mathcal{O}_N\}$, abbreviated as nodes $\mathcal{O}_i$, retrieve the required data from data sources $\{\mathcal{D}_1,...,\mathcal{D}_M\}$ (multiple exchange rate APIs), where data sources are denoted as $\mathcal{D}_j$. Different schemes have different request strategies. For example, DAON \cite{dong2023daon} and DecenTruth \cite{xiao2023decentralized} etc. MDS schemes require each node to traverse $\{\mathcal{D}_1,...,\mathcal{D}_M\}$ to obtain data. To reduce the cost of data being uploaded to the blockchain, the current method of aggregating data off-chain is mostly used, and only the final result is uploaded. Among them, the aggregation method based on threshold signature is the most common \cite{dos,du2022novel}. The oracle contract then verifies the aggregated result from the oracle network and allocates rewards to the $t$ successful nodes to incentivize honest execution of tasks. Finally, the oracle contract calls a callback interface to return the data to the user contract, completing the task.

To address the two key challenges mentioned above, this paper proposes several improvements to off-chain oracle networks. Specifically, we introduce a novel off-chain data aggregation method, TBLS, which balances distributed trust with data source diversity. Additionally, building on TBLS, we enhance the user request strategy such that a single node only needs to access one data source $\mathcal{D}_j$, while still ensuring the diversity of the final aggregated data.

\begin{figure}[hbtp]
  \centering
  \includegraphics[width=\linewidth]{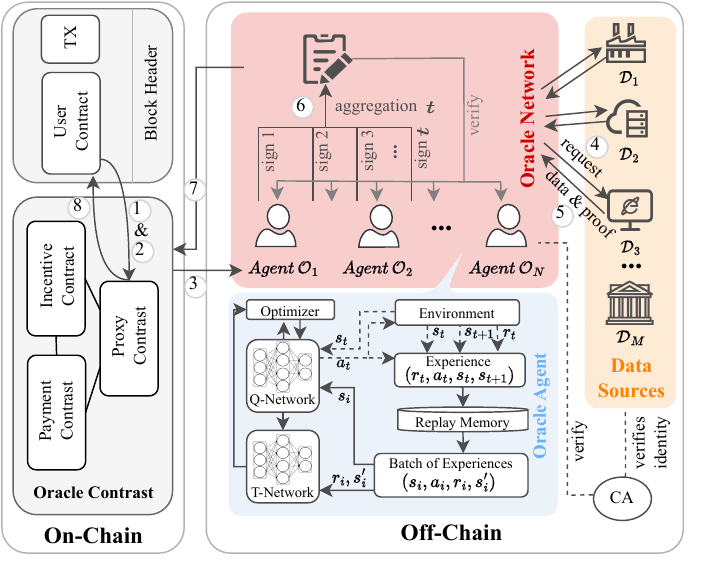}
  \caption{{\sysname} Overview.}
  \label{fig:process}
\end{figure}

The process flow of the proposed scheme is illustrated in Fig. \ref{fig:process}. The detailed steps are as follows:

\begin{enumerate}[label=\textcircled{\arabic*}]
\item The user contract calls the oracle contract interface to initiate a task request $\mathcal{Q}$, which includes the task ID $\mathcal{I}$, the requested data source set $\mathcal{D}$ and their CA certificate set $\mathcal{C}$. This interaction is performed through the proxy contract within the oracle contract, and tokens are staked and locked in the payment contract.
\item The oracle contract generates an oracle event based on the request event $\mathcal{Q}$, which is recorded on the blockchain.
\item Node $\mathcal{O}_i$ continuously listens for external data request events on the blockchain.
\item Node $\mathcal{O}_i$, acting as an agent in reinforcement learning, uses a DDQN-based reinforcement learning strategy \cite{van2016deep} to select the optimal data source $\mathcal{D}_j$ from the requested data source set $\mathcal{D}$, such that the data it reports is most likely to be aggregated by the TBLS, thereby earning rewards. This approach enhances system efficiency while ensuring data source diversity (detailed in Section \ref{rl}).
\item Node $\mathcal{O}_i$ generates proof of data source $\mathcal{P}_{i}^{j}$ based on the session and signs the retrieved data $d_j$ with $\sigma_{i}^{j} = sign(d_j, sk_{i})$.
\item Node $\mathcal{O}_i$ broadcasts and verifies $(\mathcal{I}, \{ \mathcal{P}_{i}^{j} \}, \sigma_{i}^{j})$ using the TBLS protocol. The final aggregation result is generated based on diversity requirements $K$ and the threshold $t$ (detailed in Section \ref{tbls}).
\item The first node $\mathcal{O}_i$ to successfully aggregate the result uploads the final aggregated result $\sigma$ and the original data $d$ to the blockchain for verification. The incentive contract then rewards or penalizes the oracle nodes based on their behavior, and the payment contract transfers tokens to the successful oracle nodes as compensation.
\item Finally, the oracle contract calls the callback interface to return the data to the user contract.
\end{enumerate}

\paragraph{\textbf{Security Assumptions}}\label{security} Similar to the DAON \cite{dong2023daon} and DecenTruth \cite{xiao2023decentralized}, we assume that the majority of nodes and data sources are honest. Specifically, malicious actors can compromise at most $P(D)$ of the data sources and $P(O)$ of the nodes, where $P(D) + P(O) < \frac{N+M}{2}$, $P(D) < \frac{M}{2}$, and $t < M$. At the data level, there may be collusion between malicious data sources and nodes, leading to the return of incorrect data through coordinated efforts.

\section{DETAIL DESIGN}
\label{detailed}
\subsection{TBLS Protocols}
\label{tbls}

\begin{figure}[hbtp]
  \centering
  \includegraphics[width=\linewidth]{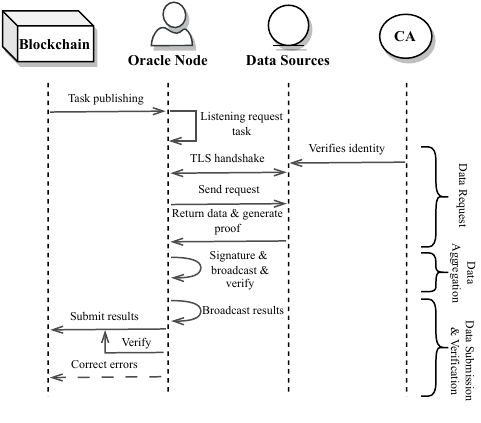}
  \caption{TBLS protocols.}
  \label{fig:TBLS}
\end{figure}

To address challenge 1, we designed a data aggregation protocol TBLS, which ensures data source diversity with low resource overhead, building on the foundation of threshold signature data aggregation methods. The detailed process is shown in Fig. \ref{fig:TBLS}.

\paragraph{\textbf{Data Request Phase}} Node $\mathcal{O}_i$ selects a data source $\mathcal{D}_j$ either randomly or strategically based on the data request task $\mathcal{Q}$ published on the blockchain. The communication between $\mathcal{O}_i$ and $\mathcal{D}_j$ is conducted using the TLS-N protocol, which allows $\mathcal{O}_i$ to generate non-repudiable proof $\mathcal{P}_{i}^{j}$ for the data source of the session, enabling third parties to verify that the data indeed originates from a specific data source \cite{ritzdorf2018tls,signingtls}. Specifically, $\mathcal{O}_i$ first establishes a TLS connection with the data source and negotiates TLS-N parameters during the handshake process. Then, $\mathcal{O}_i$ utilizes the TLS-N protocol to sign the entire session content, including handshake data and session keys. After the data transmission is completed, it generates non-repudiable proof $\mathcal{P}_{i}^{j}$ for the data source and signs the obtained data $d_j$ by its private key $sk_i$, resulting in $\sigma_{i}^{j} = sign(d_j, sk_{i})$.

\paragraph{\textbf{Data Aggregation Phase}} Node $\mathcal{O}_i$ broadcasts the data and its proof $(\mathcal{I}, \mathcal{P}_{i}^{j}, \sigma_{i}^{j})$ and also receives and verifies data and proofs from other nodes. During the aggregation process, the system must verify whether the group signature $\sigma$ from $t$ nodes meets the threshold requirements for successful aggregation, as well as whether the associated data source proofs $\left\{ \mathcal{P}_{i}^{j} \mid i \in N, j \in M \right\}$ meet the diversity requirement $K$. The parameter $K$ represents the number of data sources, typically requiring that $\frac{M}{2} < K \leq M$ to ensure data diversity and reliability. Additionally, to reduce resource overhead and communication load in the oracle network, each node is restricted from uploading data for aggregation more than once per task. If multiple uploads are detected, the node will be disqualified from participating in the current round.

\paragraph{\textbf{Data Submission and Verification Phase}} Once data aggregation and source diversity validation are successful, node $\mathcal{O}_i$ submits the group signature $\sigma$ to the blockchain and broadcasts $(\mathcal{I}, { \mathcal{P}_{i}^{j} }, \sigma)$. Similar to BLS \cite{das2024adaptively}, once other nodes verify the submission, the task is considered complete, and no further processing is required \cite{dos}. However, if other nodes fail to verify the submission or detect discrepancies between the aggregated result and the submitted result, a data correction request can be initiated. In such cases, the node will upload $(\mathcal{I}, { \mathcal{P}_{i}^{j} }, \sigma, d)$ to revise the previous answer, and the incorrect submitter will be penalized. The oracle contract will verify the correctness of the data not only based on the group public key $G_{pk}$ registered on the blockchain: $Verify_{acc}(\sigma,G_{pk},d)$, but also validate its diversity through $Verify_{div}({ \mathcal{P}_{i}^{j} }, \mathcal{C},K)$.

\subsection{Data Source Selection Strategy}
\label{rl}

Under the constraints of the TBLS aggregation protocol, nodes must carefully select data sources to maximize their chances of successful aggregation and receiving rewards. They must not only consider the response speed of the data source, aiming to be among the $t$ nodes aggregated but also take into account the selection of other nodes. For example, while $\mathcal{O}_1$ may retrieve data from $\mathcal{D}_1$ the fastest, if node $\mathcal{O}_2$ retrieves data even quicker, $\mathcal{O}_1$'s chance of earning a reward diminishes. Moreover, selecting a malicious data source reduces the likelihood of successful aggregation. Thus, nodes must selection a data source that is most likely to be successfully aggregated without knowing the selections of other nodes. This decision balances maximizing the node's benefits by ensuring both data source diversity and optimal response speed.

\paragraph{\textbf{Bayesian Game Model}} 
Since a node cannot fully grasp the information of other nodes' selection when selecting a data source, we construct the node's data source selection problem as a Bayesian game model to solve challenge 2. The game can be constructed as a quintuple of the form $G=(N, A, \Theta, P, U)$.

\begin{itemize}
    \item Set of participants \(N\) represents the set of oracle nodes participating in the game, i.e. \(N=\{\mathcal{O}_1, \mathcal{O}_2,\dots, \mathcal{O}_N\}\), where \(N\) is the total number of nodes and each \(\mathcal{O}_i\) represents an individual oracle node.
    \item Strategy Space \(A\) is the set of strategies (data sources) that each participant can select from, i.e. \(A = \{a_1,\dots,a_M\}\). Where \(M\) is the total number of data sources, and \(a_j\) represents an alternative data source \(\mathcal{D}_j\).
    \item Type set \(\Theta\) represents the possible types of each node, with the size of the set corresponding to the total number of data sources, \(M\). Formally, \(\Theta = \{\theta_1, \theta_2, \dots, \theta_M\}\), where each type \(\theta_i \in \Theta\) indicates a specific advantage a node may have when accessing data source \(\mathcal{D}_j\). This advantage could arise from factors such as the node \(\mathcal{O}_i\) being geographically closer to \(\mathcal{D}_j\), having a faster network connection, or the likelihood of other nodes accessing the same data source is lower, etc.
    \item Type probability distribution \(P\) describes the prior probability distribution of each node's type. Specifically, \(P_i(\theta_j)\) represents the probability that node \(\mathcal{O}_i\) belongs to type \(\theta_j\), based on public prior knowledge and without access to node \(\mathcal{O}_i\)'s private information (such as response time or historical selections). This probability distribution also referred to as prior belief, reflects the initial estimate of the node’s likely type.
    \item Utility function \(U\) represents the benefit or utility of each node given a particular type and strategy selection. For each node \(\mathcal{O}_i\) and type \(\theta_i\), the utility function \(U_i(a_i, \theta_i, \theta_{-i}, a_{-i})\) describes how well the node \(\mathcal{O}_i\) is able to select the strategy \(a_i\) given its type \(\ theta_i\), other nodes' types \(\theta_{-i}\) and other nodes' strategies \(a_{-i}\) when selecting its strategy \(a_i\). Here \(a_i \in A\) is the strategy of the node \(\mathcal{O}_i\), and \(a_{-i}\) is the combination of the strategies of other nodes except \(\mathcal{O}_i\).
\end{itemize}

\paragraph{\textbf{Markov Decision Process}} 
Due to the dynamic belief updating and incomplete information in the selection game of oracle nodes, the nodes must gradually learn the optimal strategy based on limited observation data, and traditional game theory makes it difficult to deal with such complex interactions \cite{fudenberg1998theory}. Reinforcement Learning (RL) is suitable for approximating the optimal strategy through continuous trial and error in a dynamic environment, and thus we can convert the problem into a Partially Observable Markov Decision Process (POMDP) \cite{liang2021gadqn,morato2023inference}. Further, to optimize the solution process, we simplify it to a Markov Decision Process (MDP), denoted as $(S, A, T, R,\gamma)$.

To transition from partial to full observability, we incorporate type inference, or beliefs, from Bayesian games into the state \(S\). Specifically, we design an advantage matrix as state \(S\) to represent both observable and unobservable information uniformly. During the construction of the advantage matrix, nodes can continuously update the belief $P_{-i}(\theta)$ according to the strategy they select ( unobservable information ) and the received aggregation success results ( observable information ), to improve the accurate inference of other node types.

 \begin{figure}[hbtp]
  \centering
  \includegraphics[width=\linewidth]{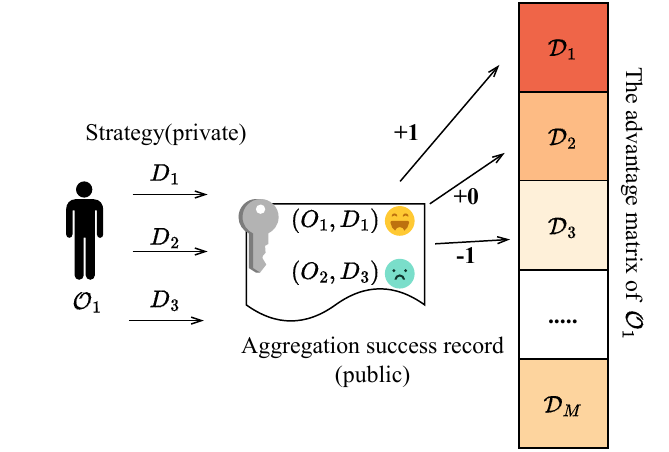}
  \caption{The influence of different strategies of node $\mathcal{O}_1$ on the advantage matrix.}
  \label{fig:matrix}
\end{figure}

Each node maintains its advantage matrix \(\mathbb{T}\), which tracks its performance across various data sources. By updating this matrix, nodes adjust their belief about the types of other nodes after each task, based on the aggregation results. As illustrated in Fig. \ref{fig:matrix}, when a node successfully aggregates data from a particular source, the corresponding entry in the advantage matrix increases by \(+1\), indicating that the node holds an advantage for that data source. Conversely, if the node selects the same data source as others but fails to aggregate, the entry decreases by \(-1\), suggesting a lack of advantage. If no other node successfully aggregates data from the select source, no advantage is inferred, and the entry remains \(0\). This mechanism allows the advantage matrix to reflect the node’s private selections and the publicly available aggregation results.

The set of actions $A$ is the same as the strategy space $A$ of the Bayesian game. The state transfer function $T(s'|s,a)$ denotes the probability that the system transfers to the next state $s'$ after selecting action $a$ in the current state $s$. The $\gamma \in (0,1]$ is a discount factor that represents the current value of the future reward.

The reward $R$ corresponds to the utility function $U$. In this task, $R$ denotes the reward obtained by the node after successful aggregation. The $t$ nodes that are successful in aggregation in each task will receive the reward, while the other nodes cannot receive the reward and are agnostic about their data source selection. This reward structure encourages nodes to select the appropriate data source to increase the aggregation success rate and maximize the reward.

\begin{equation}
\label{reward}
\left.R=\left\{\begin{matrix}{1,}&{\mathrm{reach~consensus,}}\\{0,}&{\mathrm{otherwise.}}\\\end{matrix}\right.\right.
\end{equation}

\paragraph{\textbf{Strategy Solving}}
In the given Markov Decision Process (MDP) \((S, A, T, R, \gamma)\), we employ the Double Deep Q-Network (DDQN) algorithm \cite{van2016deep} to find the optimal action decision policy \(\pi^*\). DDQN aims to learn the state-action value function \(Q(s, a)\), which satisfies the Bellman equation:

\begin{equation}
Q(s, a) = \mathbb{E}[R_t + \gamma \max_{a'} Q(s_{t+1}, a') | s_t = s, a_t = a].
\end{equation}

The goal of DDQN is to approximate the solution to this equation, leading to the optimal policy \(\pi^*\).

The detailed steps are as follows:
\ding{182} Initialize the main Q-network \(Q_{\theta}\) and the target Q-network \(Q_{\theta^-}\) with parameters \(\theta\) and \(\theta^-\), respectively. Also, set the experience replay buffer \(\mathcal{D}\), exploration rate \(\epsilon\), and discount factor \(\gamma\), along with other hyperparameters. \ding{183} At each time step \(t\), in the current state \(s_t\), select an action \(a_t\) using an \(\epsilon\)-greedy policy:

\begin{equation}
       a_t = 
   \begin{cases} 
   \text{random action,} & \text{with probability} \epsilon , \\ 
   \arg\max_a Q_{\theta}(s_t, a) ,& \text{with probability } 1 - \epsilon .
   \end{cases}
\end{equation}

\ding{184} Execute the selected action \(a_t\) in the environment, receiving a reward \(r_t\) and observing the next state \(s_{t+1}\). \ding{185}  Store the experience tuple \((s_t, a_t, r_t, s_{t+1})\) into the experience replay buffer \(\mathcal{D}\). \ding{186} Randomly sample a batch of experience \((s_i, a_i, r_i, s_{i+1})\) with size \(B\) from the buffer \(\mathcal{D}\), and calculate the target Q-value \( y_i \).

\begin{equation}
    y_i = r_i + \gamma Q_{\theta^-}(s_{i+1}, \arg\max_a Q_{\theta}(s_{i+1}, a)).
\end{equation}

 Here, the target Q-value depends on the estimation of the target network \(Q_{\theta^-}\) to mitigate the problem of overestimation. According to the Bellman equation, \( y_i \) is an approximation of \( Q(s_i, a_i) \).

\ding{187} Update the main Q-network parameters using gradient descent and Mean Squared Error (MSE) loss function, adjusting the parameters \(\theta\) of the main Q-network.
\begin{equation}
    L(\theta) = \frac{1}{B} \sum_{i=1}^{B} (y_i - Q_{\theta}(s_i, a_i))^2 .
\end{equation}

\ding{188} Periodically copy the main Q-network parameters \(\theta\) to the target network \(\theta^-\), ensuring training stability.

By repeating steps \ding{183} to \ding{188}, DDQN progressively updates the Q-values and eventually converges to an approximation of the state-action value function \(Q(s, a)\). The final optimal policy is obtained by selecting the action with the maximum Q-value for each state:
\begin{equation}
    \pi^*(s) = \arg\max_a Q(s, a).
\end{equation}

\section{EXPERIMENT AND SAFETY ANALYSIS}
\label{experiment}

We implemented the proposed solution using Python 3.9 to simulate the oracle network. Specifically, we constructed the oracle network comprising 50 oracle nodes and 20 data sources. The oracle nodes were deployed on a platform equipped with an Intel(R) Core(TM) i7-9700F processor and 16 GB of RAM. The parameter settings are presented in Table \ref{tab:Experimental-settings}. To fully demonstrate the effectiveness of {\sysname}, we compared it against the following four baselines:
\begin{itemize}
    \item DAON \cite{dong2023daon}. The representative of the MDS oracle scheme guarantees the diversity and reliability of the data by requiring each node to traversal access all \(M\) data sources. 
    \item IoT \cite{xian2024distributed}. The performance improvement scheme of MDS oracle uses the reputation mechanism to allocate $\mathcal{K} = 2$ nodes with fast response and high success rates for different data sources.
    \item Simple. Each node simply selects a data source randomly for access.
    \item Simple-n. Each node simply and randomly selects $n=2$ data sources to access. (Default Simple-2)
\end{itemize}

\begin{table}[h]
\centering
\caption{Experimental parameter setting}
\begin{tabular}{lc}
\hline
Parameter name & Value  \\ \hline
total oracle nodes $N$ & 50  \\
total data sources $M$ & 20 \\
signature threshold $t$ & 20  \\ 
diversity requirements $K$  & 18  \\ 
oracle task counts & 1000 \\
network latency distribution (s) & $U(0.1, 2.3)$  \\
learning rate $\eta$ & 5e-4 \\
 exploration rate $\epsilon$ & 0.05 \\
discount factor $\gamma$ & 0.99 \\
memory size  & 1000 \\ \hline
\end{tabular}
\label{tab:Experimental-settings}
\end{table}

\begin{figure*}[t!]
\centering
\subfloat[Response Time]{
    \includegraphics[width=0.23\linewidth]{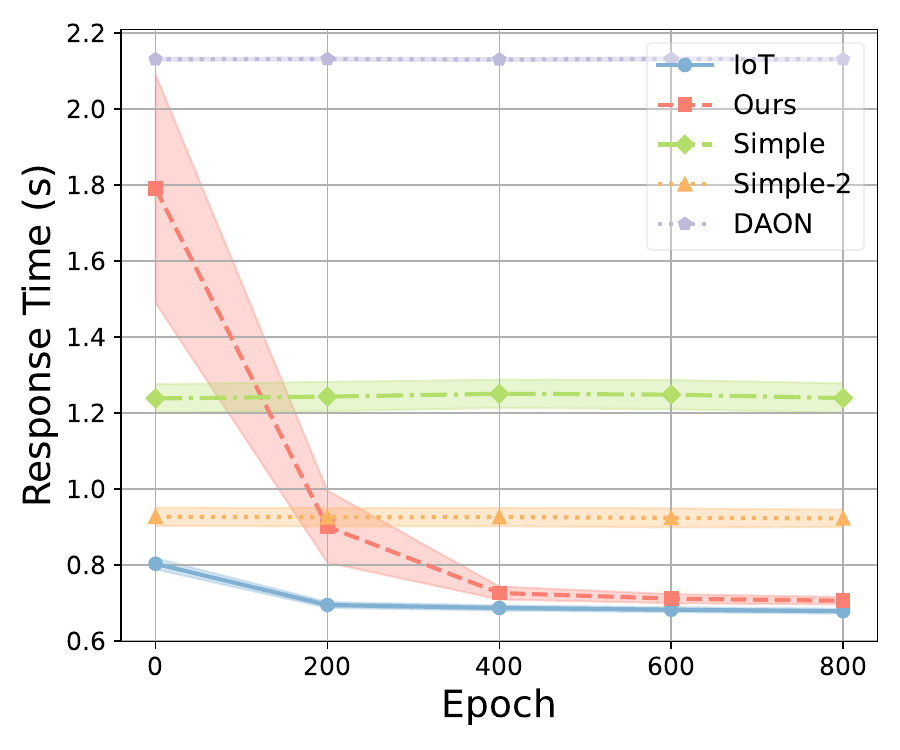}
    \label{subfig:all_time}
}
\hfill
\subfloat[Diversity of Data Sources]{
    \includegraphics[width=0.23\linewidth]{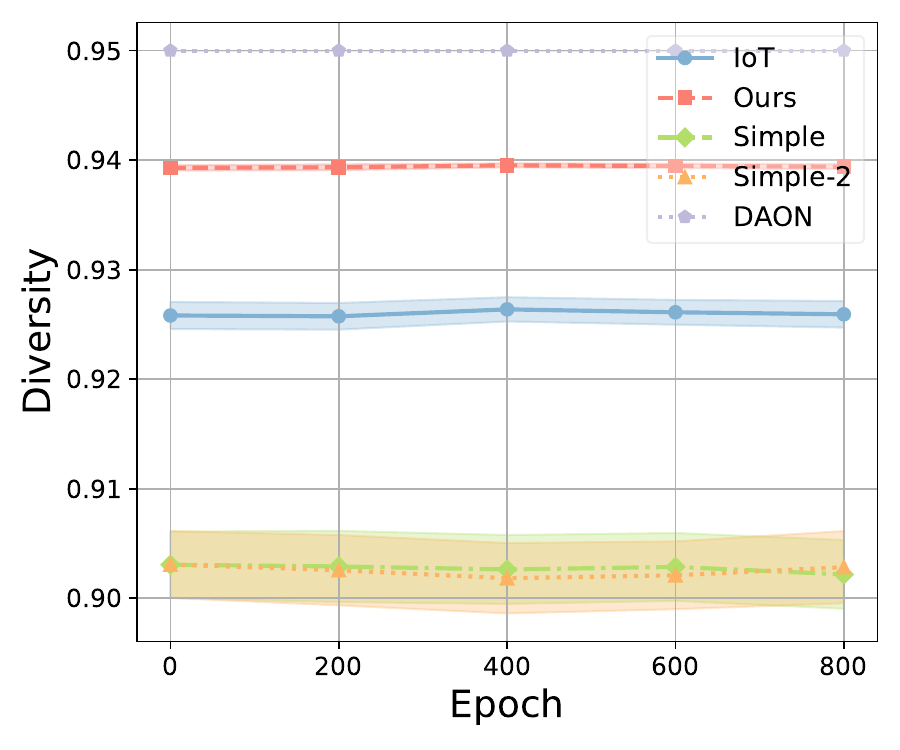}
    \label{subfig:all_div}
    }
\hfill
\subfloat[Loss]{
    \includegraphics[width=0.23\linewidth]{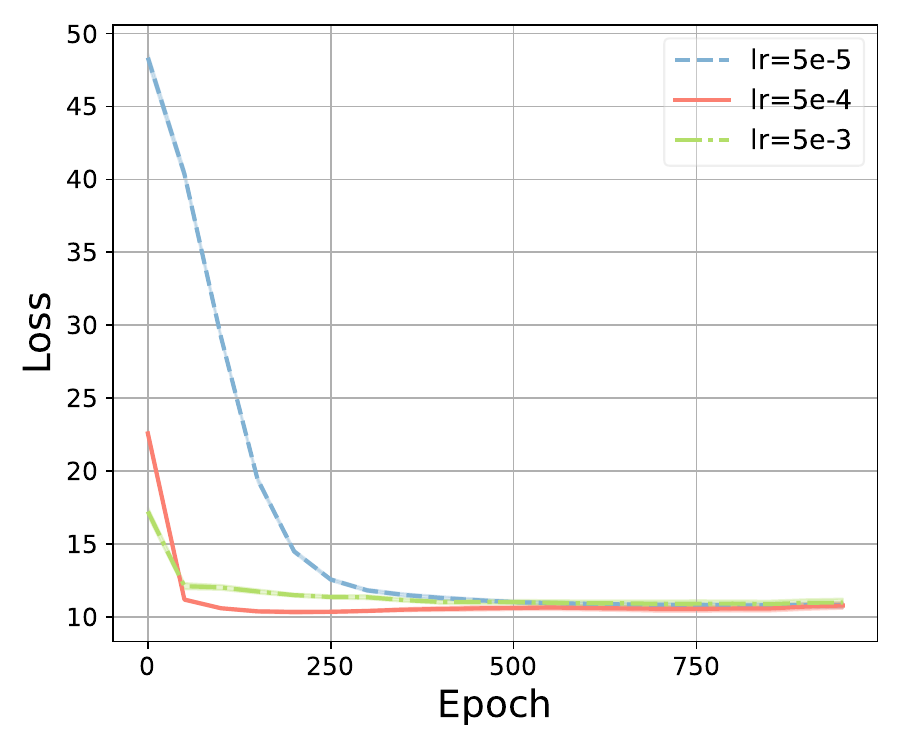}
    \label{subfig:rl_loss}
    }
\hfill
\subfloat[Reward]{
    \includegraphics[width=0.23\linewidth]{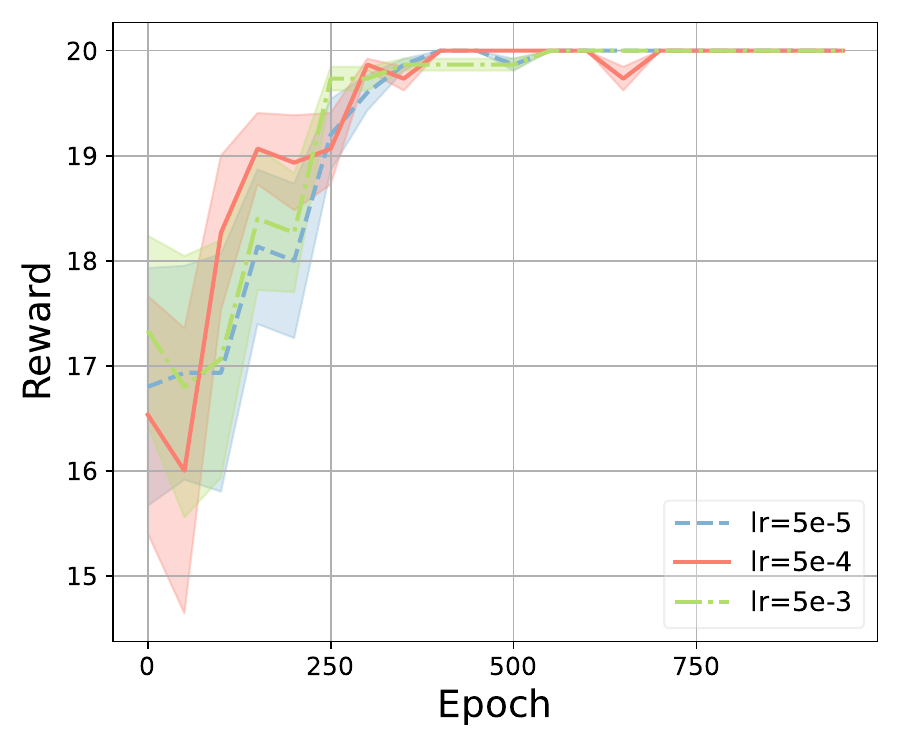}
    \label{subfig:rl_reward}
    }
\caption{Baseline comparison, diversity and response time, effectiveness of proposed reinforcement learning algorithms.}
\label{fig:all_vs}
\end{figure*}

\subsection{Baseline Comparison}
We measure the diversity of data sources using a variant of the information entropy formula, $D = 1 - \sum_{i=1}^{n} p_i^2$. where \(p_i\) is the frequency of occurrence of the category \(i\). The value of $D$ ranges from \(D \in [0, 1 - \frac{1}{n}]\), $n$ is the total number of categories in the final aggregated data. When the frequencies of all categories are uniformly distributed, \(D\) will approach the maximum value \(1 - \frac{1}{n}\). In the same case, the value of \(D\) is 0.

Fig. \ref{subfig:all_time} - Fig.\ref{subfig:all_div} illustrate the node response time and data source diversity under the experimental setup for different approaches. It is evident that in the Simple and IoT schemes, nodes do not need to aggregate data from all sources, which leads to a significant advantage in response time. However, this comes at the cost of reduced data source diversity. In particular, the IoT approach achieves excellent performance by using reputation contracts to assign nodes with fast response times for each data source. However, its major drawback is the reliance on a TEE to ensure the credibility of the node response times, as well as the substantial storage and computation overhead on the blockchain to calculate node reputations. 
In contrast, {\sysname} achieves comparable response times to IoT without the need for TEE support or increased on-chain computation and storage costs, while also ensuring higher data source diversity. This demonstrates that the proposed scheme effectively addresses Challenges 1 and 2, enhancing system efficiency while maintaining diversity of data sources.

Fig. \ref{subfig:rl_loss}-\ref{subfig:rl_reward} shows the proposed reinforcement learning method's loss and reward curves under different learning rates. In the case of $M=20, t=20$, the rewards under different settings eventually approach 20, i.e., there are data from at least $K$ different data sources in each aggregation such that $t$ nodes are rewarded. The proposed method demonstrates strong usability and can eventually learn and approximate the optimal policy, depending on the environment, even when different learning rates are set.

Table \ref{tab:time-used} presents the average time consumption over the last 100 tasks for different schemes. Although the proposed solution introduces Reinforcement Learning (RL) to learn the optimal data source selection strategy, the design of the advantage matrix simplifies the RL training process. Nodes only require minimal computational resources to significantly reduce communication time. As a result, {\sysname} achieves response times comparable to the IoT approach without incurring significant on-chain computation and storage costs or relying on the TEE. Compared to other approaches like DAON and Simple, which also do not require TEE, {\sysname} reduces response time by 23.5\% relative to the best baseline (Simple-2).

Table \ref{tab:extr-add} shows the number of data source accesses and the additional overhead for different schemes. Compared to traditional MDS oracle solutions such as DAON and IoT, {\sysname} requires each node to only access a data source. It is particularly important for data sources that charge fees based on the number of accesses \cite{chainlink}. Furthermore, in contrast to performance-optimized solutions like IoT, {\sysname} does not introduce additional on-chain overhead or require specialized technical support, significantly reducing gas consumption and the cost of acquiring TEE devices, thereby greatly improving usability.

\begin{table}[]
\caption{Overall time consumption analysis of oracle nodes}
\resizebox{\linewidth}{!}{%
\begin{tabular}{lccccc}
\hline
 & Ours & Simple-1 & Simple-2 & IoT & DAON \\ \hline
Computational time(s) & 0.0025(±0.0003) & - & - & - & - \\
Correspondence(s) & 0.692(±0.037) &1.219(±0.210) & 0.908(±0.126) & 0.672(±0.036) & 2.111(±0.007)\\ 
\rowcolor{lightgray} Total(s) & \textbf{0.695(±0.037)} &1.219(±0.210) & 0.908(±0.126) & \textbf{0.672(±0.036)} & 2.111(±0.007)\\  \hline
\end{tabular}
}
\label{tab:time-used}
\end{table}

\begin{table}[]
\caption{Number of data source accesses and additional overhead for different schemes}
\resizebox{\linewidth}{!}{%
\begin{tabular}{lccccc}
\hline
 & \cellcolor{gray!50} Ours & Simple-1 & Simple-n & IoT & DAON \\ \hline
Number of Data Source Accesses & \cellcolor{gray!50} $1$ &  $1$ &  $n$ &  $(0 - M \times \mathcal{K})$ & $M$ \\
Additional On-chain Overhead & \cellcolor{gray!50} \ding{56} &  \ding{56} &  \ding{56} &  Node Selection \& State Store & \ding{56} \\
Additional Technical Support & \cellcolor{gray!50} \ding{56} &  \ding{56} &  \ding{56} &  TEE & \ding{56} \\
\hline
\end{tabular}
}
\label{tab:extr-add}
\end{table}

\begin{figure*}[t!]
\centering
\subfloat[Ours]{
    \includegraphics[width=0.31\linewidth]{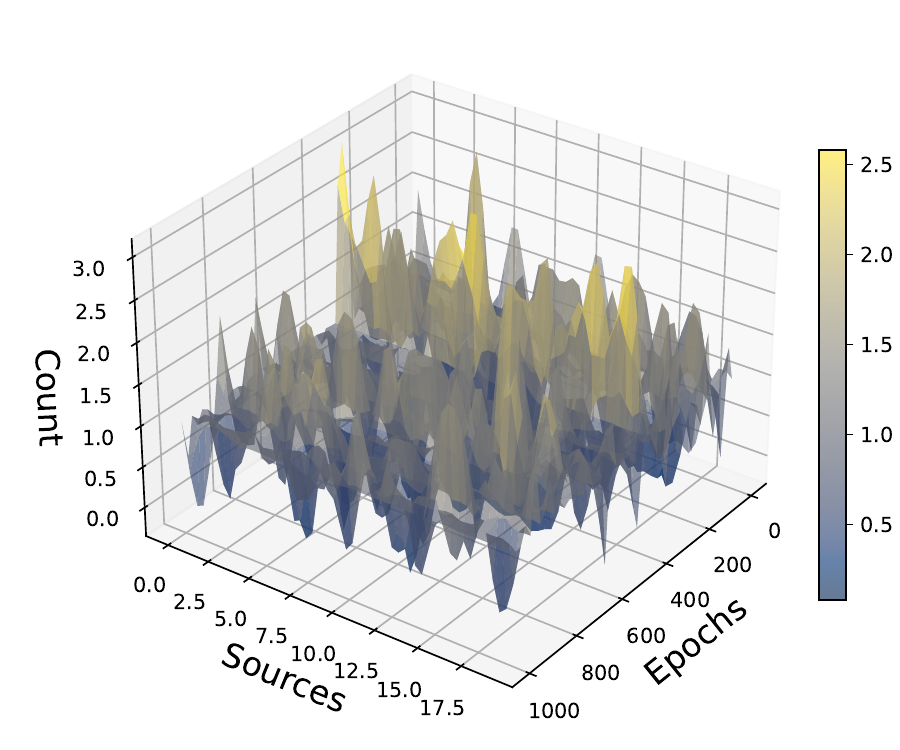}
    \label{subfig:div_ours}
}
\hfill
\subfloat[Simple]{
    \includegraphics[width=0.31\linewidth]{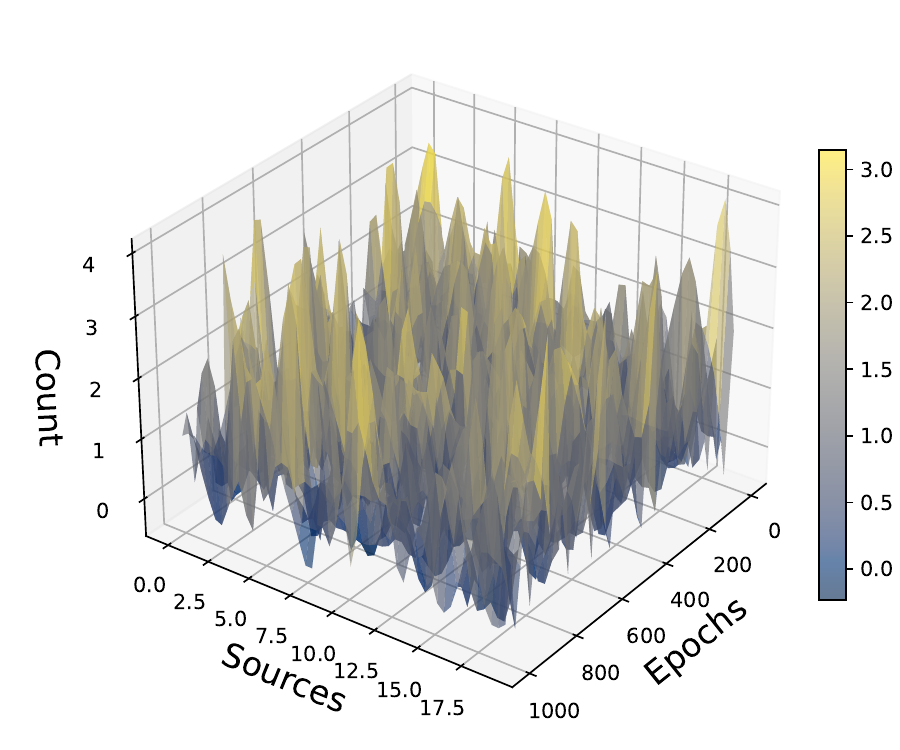}
    \label{subfig:div_simple}
    }
\hfill
\subfloat[DAON]{
    \includegraphics[width=0.31\linewidth]{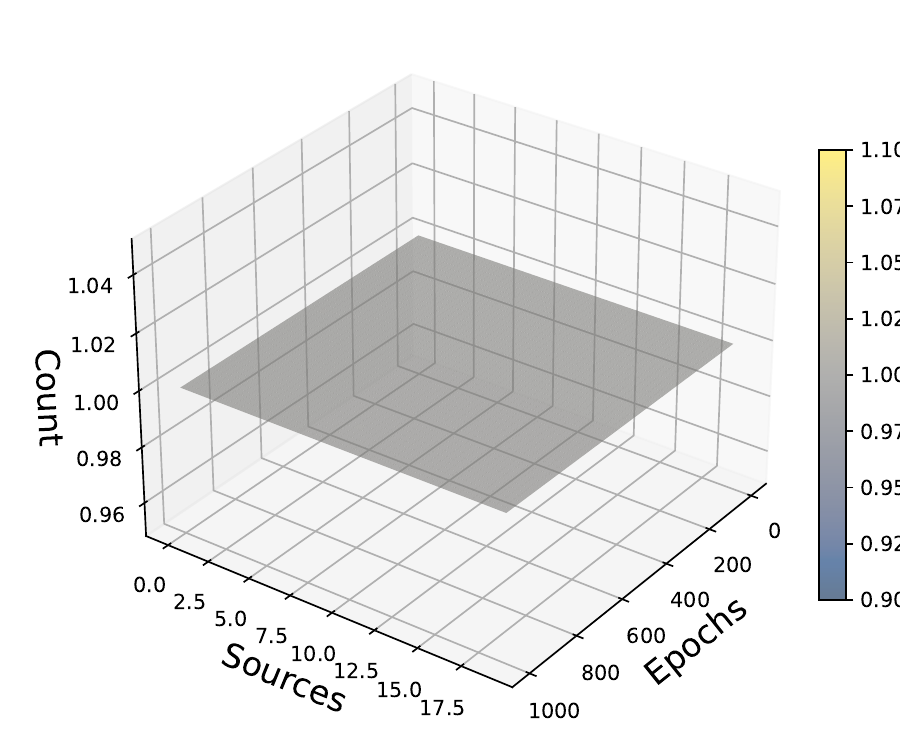}
    \label{subfig:div_daon}
    }
\hfill
\caption{The significance of diversity in the number of data sources selected for each task.}
\label{fig:divs}
\end{figure*}

\subsection{Diversity Analysis}
Fig. \ref{fig:divs} shows the distribution of the final aggregated data sources over 1000 tasks. First of all, DAON, as a representative of the traditional MDS scheme, guarantees that the final aggregated results are completely from different data sources by redundantly accessing all the data. On the other hand, {\sysname} can set $K=18$ in TBLS to constrain the source diversity of the final aggregated data. Compared with completely randomly selecting data sources without constraints, the diversity of data sources is higher and the final data sources are more balanced. It verifies the effectiveness of TBLS in solving Challenge 1. As the epoch increases,  the data source becomes more and more even, inhibiting the possibility of data being manipulated, which further confirms the effectiveness of {\sysname}.

\begin{figure*}[t]
\centering
\subfloat[Ours]{
    \includegraphics[width=0.31\linewidth]{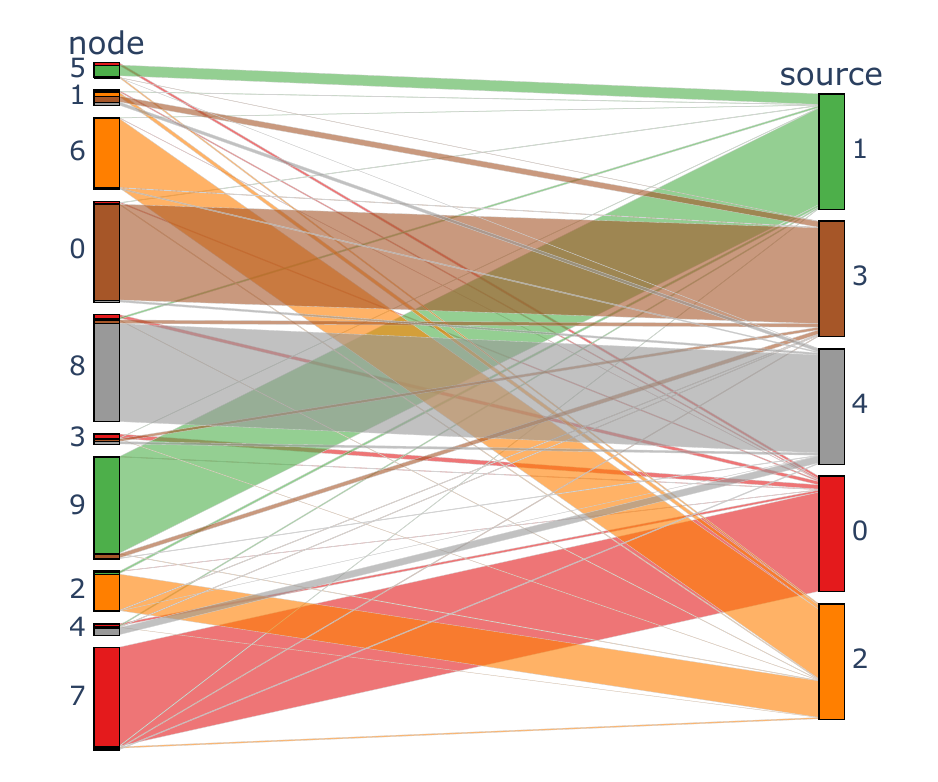}
    \label{subfig:div_ours_n2s}
}
\hfill
\subfloat[DAON]{
    \includegraphics[width=0.31\linewidth]{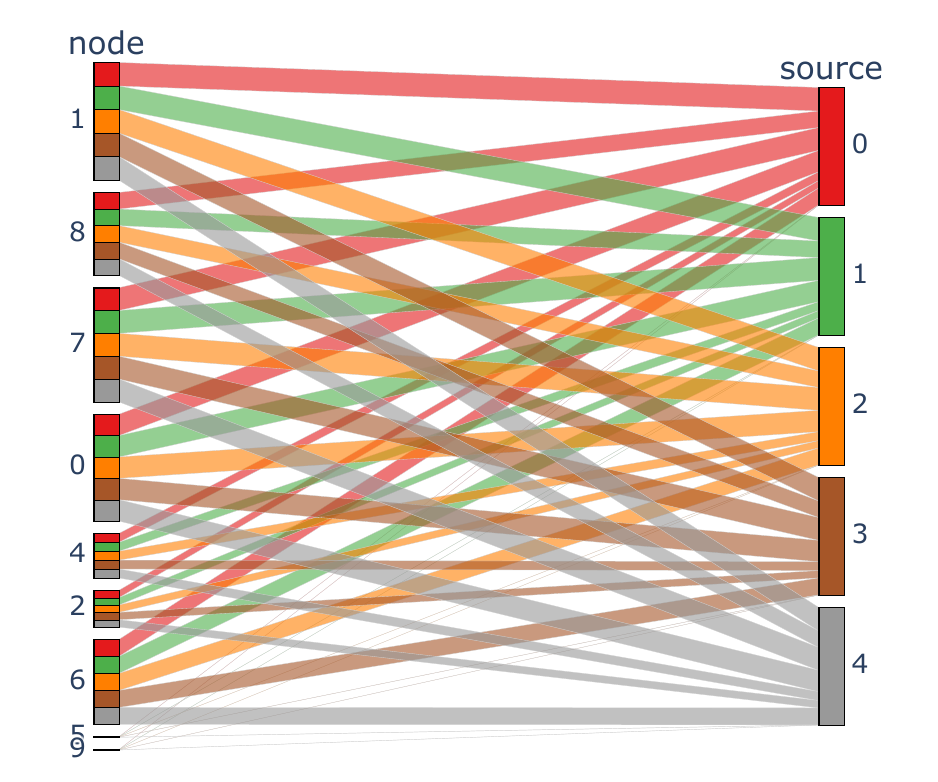}
    \label{subfig:div_daon_n2s}
    }
\hfill
\subfloat[IoT]{
    \includegraphics[width=0.31\linewidth]{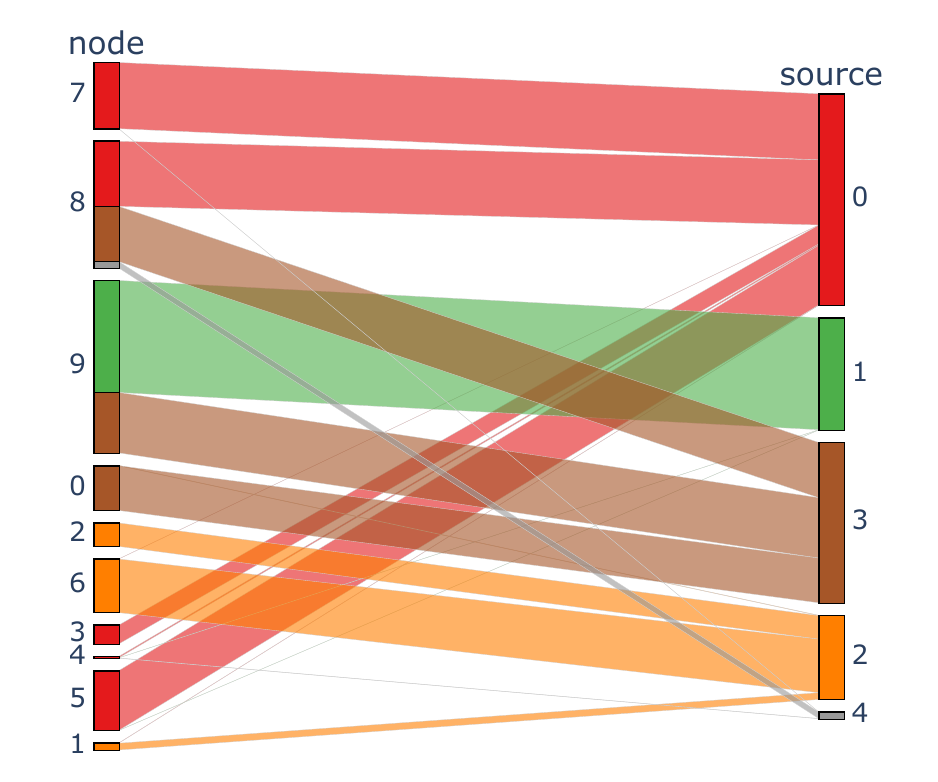}
    \label{subfig:div_iot_n2s}
    }
\hfill
 \subfloat[Simple-1]{
    \includegraphics[width=0.31\linewidth]{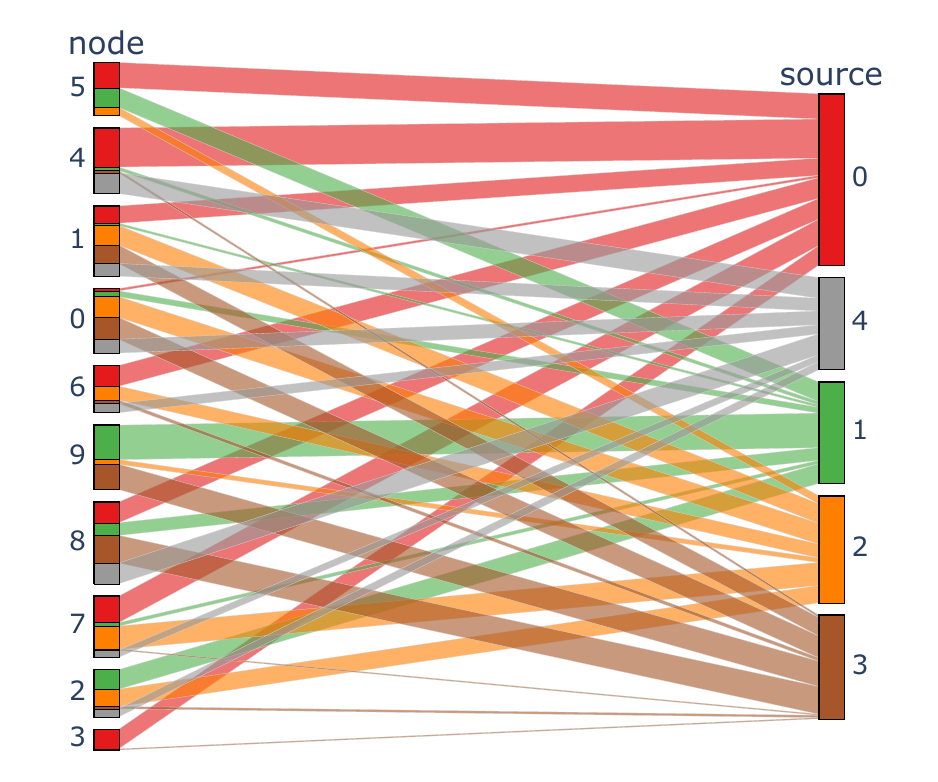}
    \label{subfig:div_simple_n2s}
    }
\hfill
 \subfloat[Simple-2]{
    \includegraphics[width=0.31\linewidth]{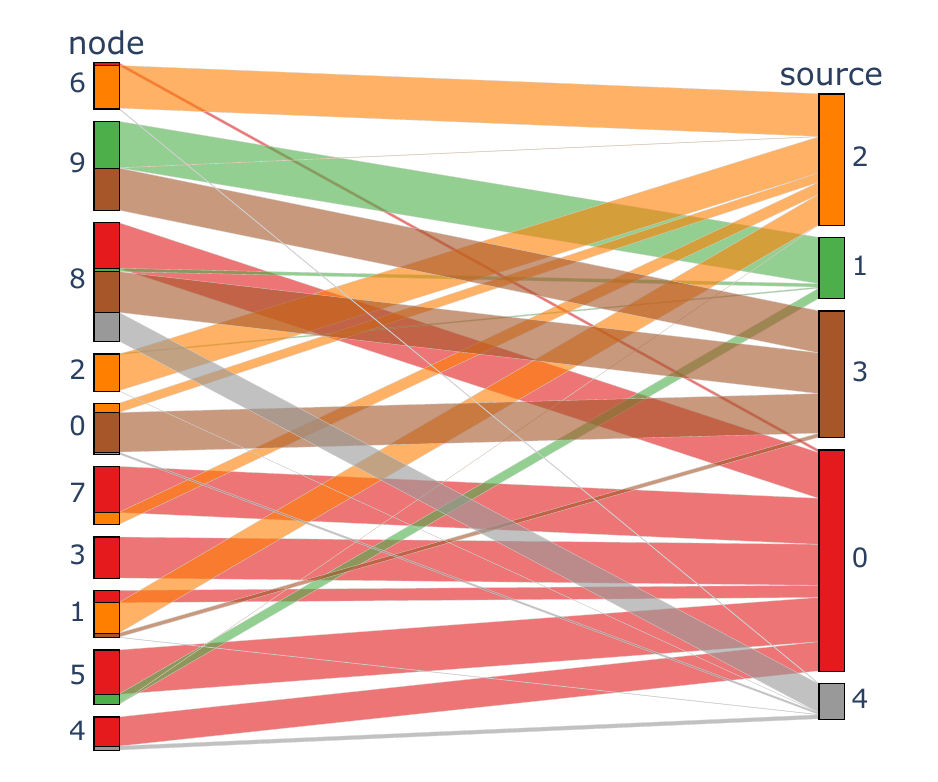}
    \label{subfig:div_simple_2_n2s}
    }
\hfill
 \subfloat[Simple-5]{
    \includegraphics[width=0.31\linewidth]{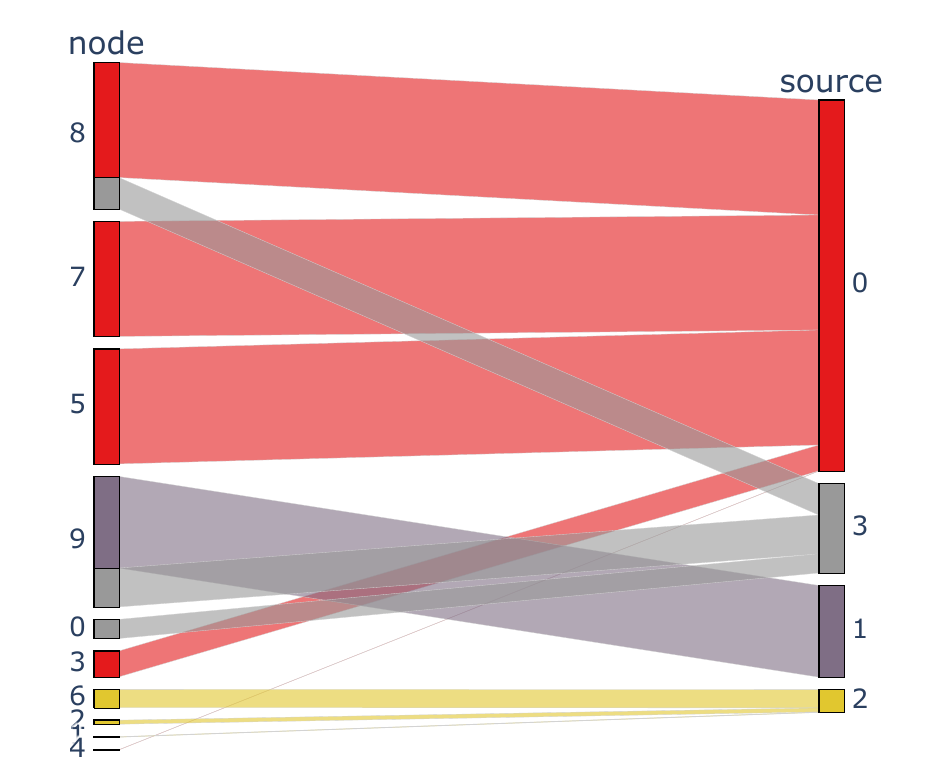}
    \label{subfig:div_simple_5_n2s}
    }
\hfill
\caption{Access of nodes to data sources in different scenarios. For example, 10 nodes access 5 data sources 1000 times.}
\label{fig:n2s}
\end{figure*}

In Fig. \ref{fig:n2s}, we further analyze the data source access patterns of different schemes. Using $N=10$, $M=5$, and $t=5$ as an example, where 10 oracle nodes perform 1000 tasks from 5 data sources, Simple-1, Simple-2, and Simple-5 represent cases where nodes randomly select \{1, 2, 5\} data sources for access, respectively. Simple-5 represents an extreme case focusing on minimizing response time, where each node accesses data from all 5 data sources and broadcasts the results for aggregation. The fastest $t=5$ responses are selected for final aggregation. 

As shown in Fig. \ref{fig:n2s}, the final aggregated data sources for schemes like IoT and Simple, which do not consider data source diversity, exhibit an uneven distribution. In particular, the Simple-5 scheme, which entirely focuses on achieving rapid responses, may produce erroneous aggregated results if data sources 0 and 3 conspire to return incorrect results, even with the use of the BLS protocol. 
In contrast to DAON, our proposed scheme does not require fetching data from all $M$ sources, significantly reducing resource overhead. Moreover, similar to IoT, our scheme identifies implicit matching relationships between node 9 and data source 1 (indicated by the green band), which greatly enhances its response speed. However, to ensure data source diversity, our approach avoids situations where a large number of nodes access data from source 0, thus mitigating the risk of aggregation errors mentioned above. The results indicate that {\sysname} not only ensures data source diversity without concentrating on specific sources but also enhances response speed by selecting sources that can provide quick responses, similar to IoT.

\begin{figure*}[htbp]
\centering
\subfloat[Malicious Nodes]{
    \includegraphics[width=0.31\linewidth]{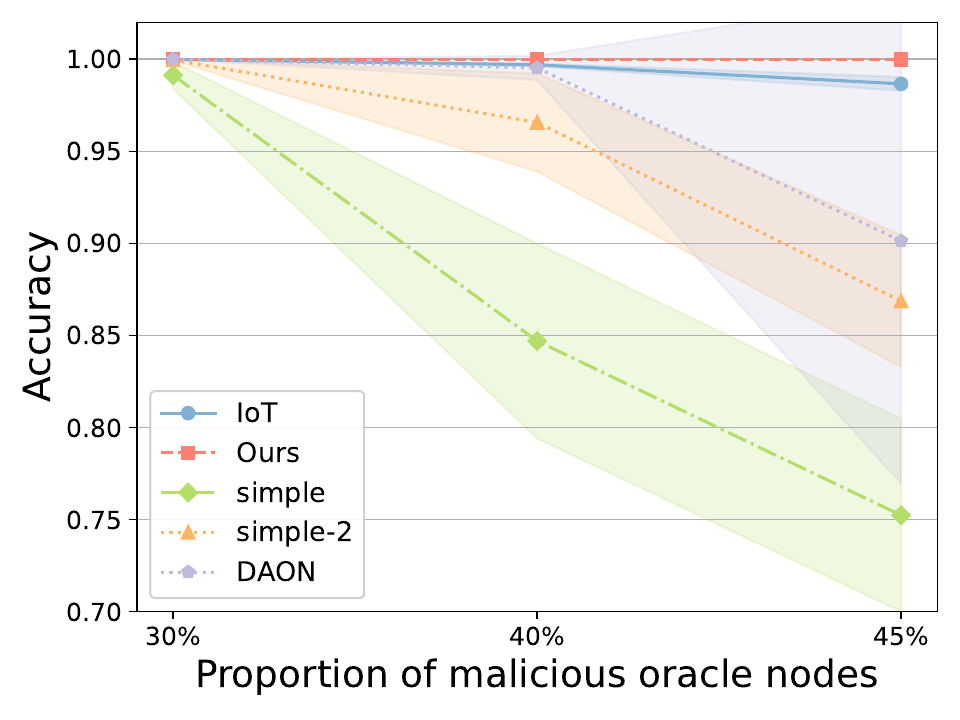}
    \label{subfig:malicious_nodes}
}
\hfill
\subfloat[Malicious Data Sources]{
    \includegraphics[width=0.31\linewidth]{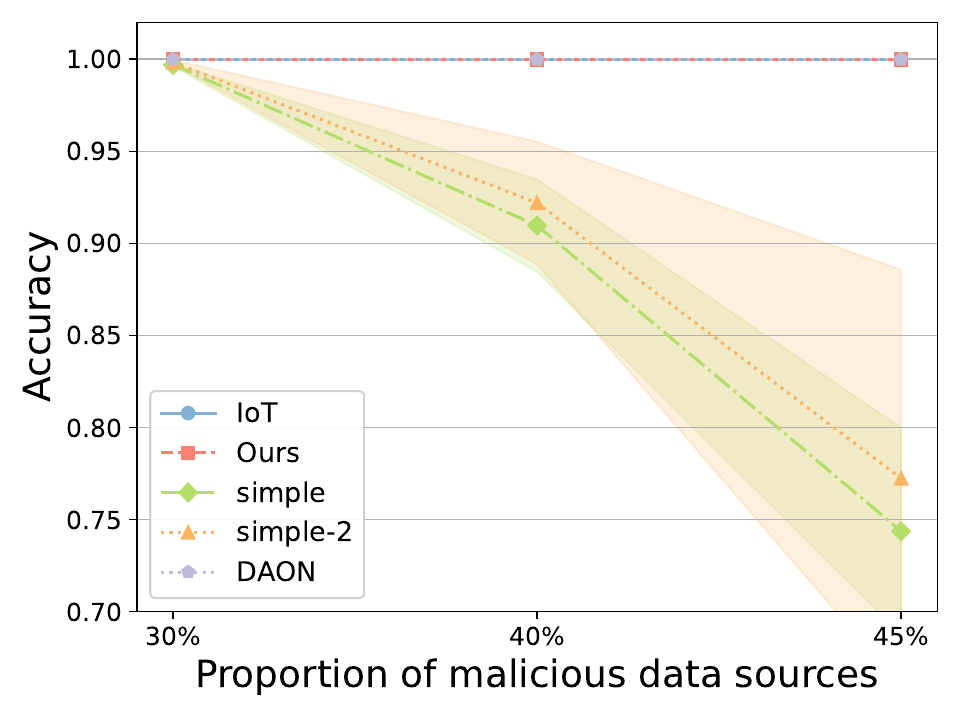}
    \label{subfig:malicious_sources}
    }
    \hfill
\subfloat[Malicious Nodes \& Data Sources]{
    \includegraphics[width=0.31\linewidth]{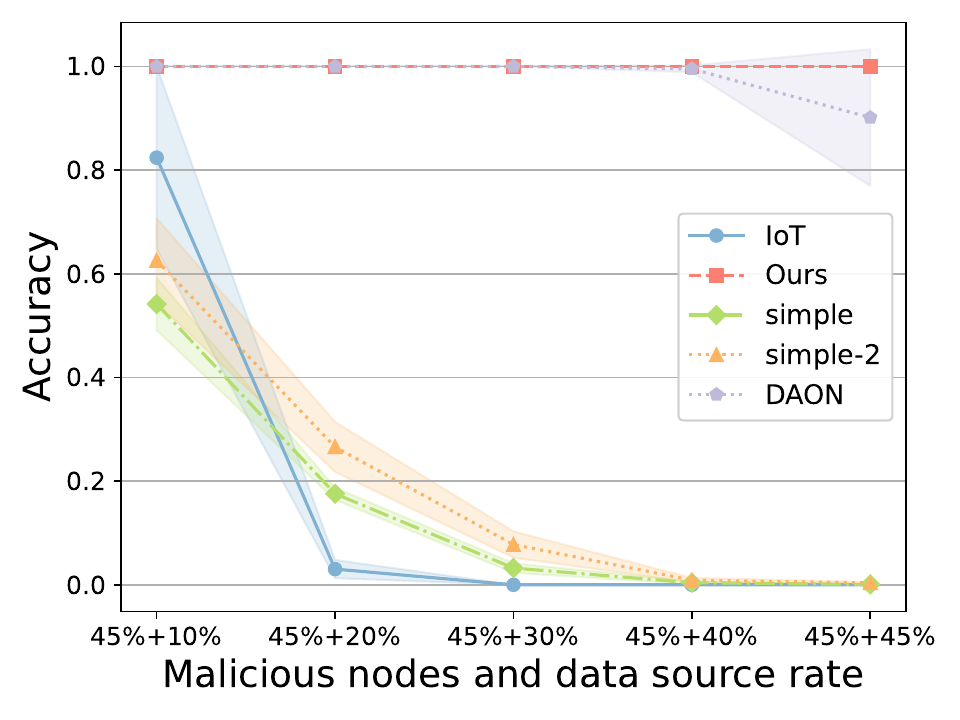}
    \label{subfig:malicious_both}
    }
\caption{The accuracy of data under varying degrees of malicious nodes and data sources.}
\label{fig:malicous}
\end{figure*}

\subsection{Attack Defence}
Fig. \ref{fig:malicous} demonstrates the ability of different schemes to defend against the malicious behavior of nodes and data sources when they conspire to return incorrect results. As Fig. \ref{subfig:malicious_nodes}, when the proportion of malicious nodes increases, the data accuracy of other baselines gradually decreases, especially when the number of malicious nodes $0.45 \times N \approx 23$ is greater than the threshold $t = 20$, the IoT and DAON methods cannot guarantee the accuracy of their data. The proposed scheme, on the other hand, can ensure the integrity and accuracy of the data uploaded by the nodes with the help of the proofs generated by the TLS interaction process.

Fig. \ref{subfig:malicious_sources} demonstrates the impact of the proportion of malicious data sources on the accuracy of the final data, as the problem of data aggregation error exposed in Fig. \ref{fig:n2s} when the proportion of malicious data sources increases, the schemes such as Simple are not able to guarantee the accuracy of the data they acquire. In contrast, {\sysname} and schemes such as DAON guarantee the accuracy of their data due to the diversity of their data sources.

Fig. \ref{subfig:malicious_both} shows the data accuracy of different schemes in the extreme case when the nodes conspire with the data source to return the incorrect data. As the proportion of malicious nodes increases, the accuracy rate of schemes such as IoT and Simple decreases rapidly. Even DAON returns incorrect data because the number of malicious nodes is larger than the threshold $t$. In contrast, the proposed scheme maintains data accuracy under the basic assumption that the total number of malicious nodes + malicious data sources is less than $\frac{N+M}{2}$.

\begin{figure*}[htbp]
\centering
\subfloat[TBLS Successful Aggregation Rate]{
    \includegraphics[width=0.31\linewidth]{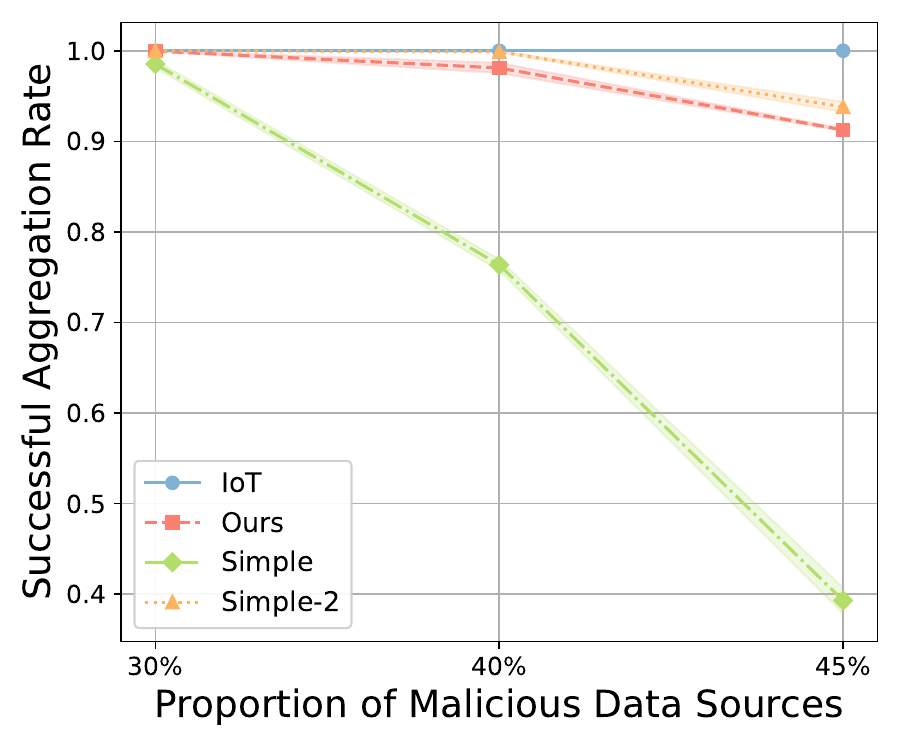}
    \label{subfig:tbls_success_rate}
}
\hfill
\subfloat[TBLS Retry Count]{
    \includegraphics[width=0.31\linewidth]{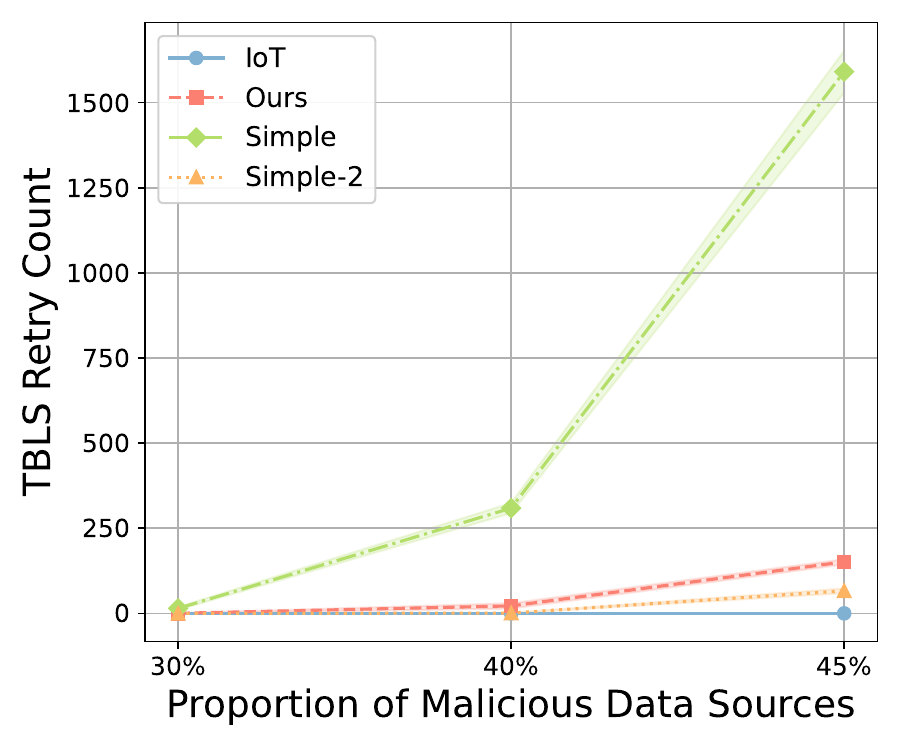}
    \label{subfig:tbls_retry_count}
    }
    \hfill
\subfloat[Response Time]{
    \includegraphics[width=0.31\linewidth]{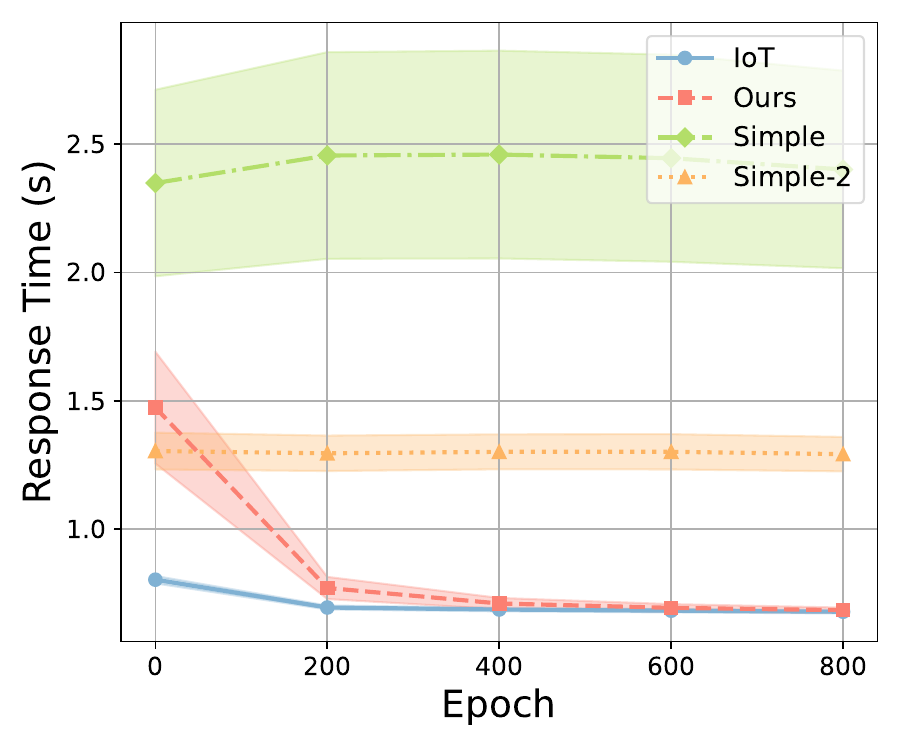}
    \label{subfig:tbls_time}
    }
\caption{Aggregation success rate, number of retries, response time when both use TBLS.}
\label{fig:tbls}
\end{figure*}

\begin{figure}[htbp]
\centering
\subfloat[TBLS Successful Aggregation Rate \& Retry Count]{
    \includegraphics[width=0.45\linewidth]{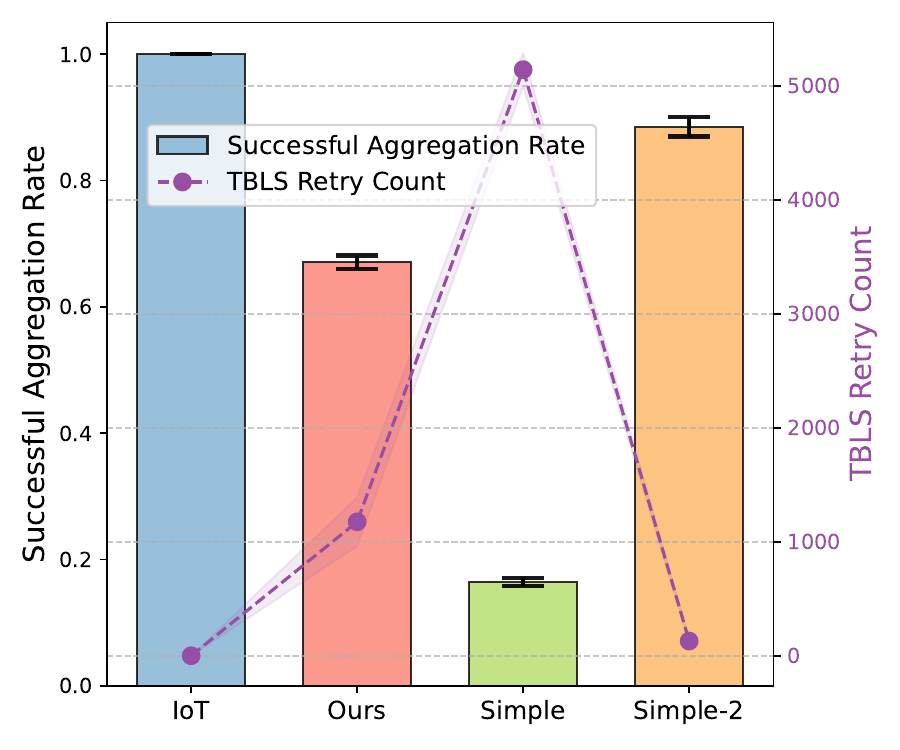}
    \label{subfig:tbls_success_rate_20}
}
\hfill
\subfloat[TBLS Retry Count]{
    \includegraphics[width=0.45\linewidth]{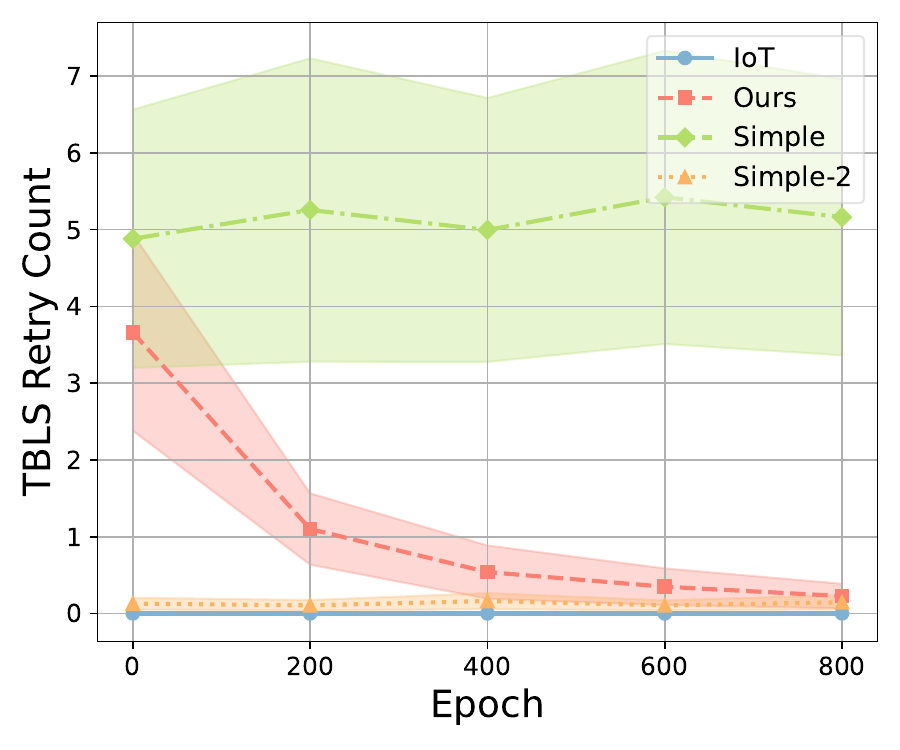}
    \label{subfig:tbls_retry_count_20}
    }
\caption{Aggregation success rate, number of retries in extreme cases with TBLS.}
\label{fig:tbls2}
\end{figure}

\begin{figure*}[htbp]
\centering
\subfloat[loss]{
\includegraphics[width=0.23\linewidth]{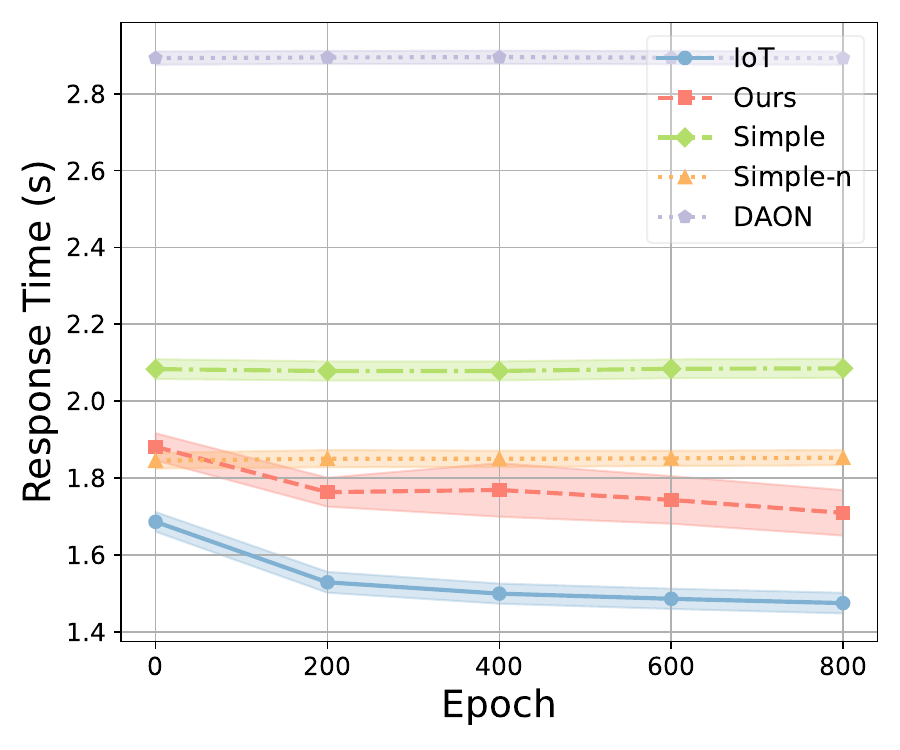}
    \label{subfig:network_time}
    }
\hfill
\subfloat[reward]{
\includegraphics[width=0.23\linewidth]{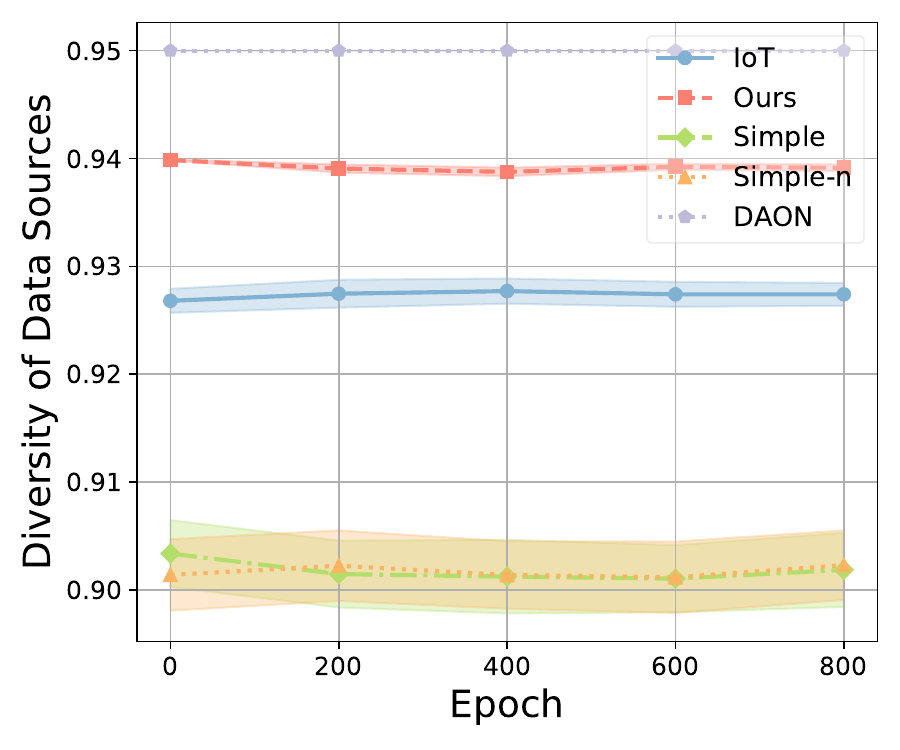}
    \label{subfig:network_div}
    }
\hfill
\subfloat[response times]{
\includegraphics[width=0.23\linewidth]{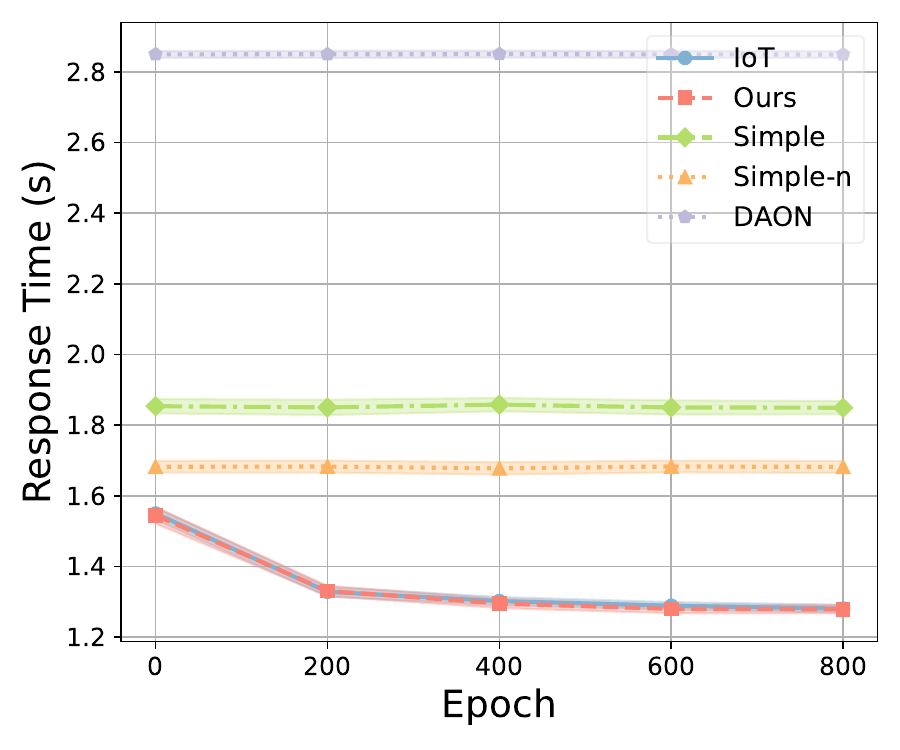}
    \label{subfig:robustness_time}
}
\hfill
\subfloat[diversities]{
\includegraphics[width=0.23\linewidth]{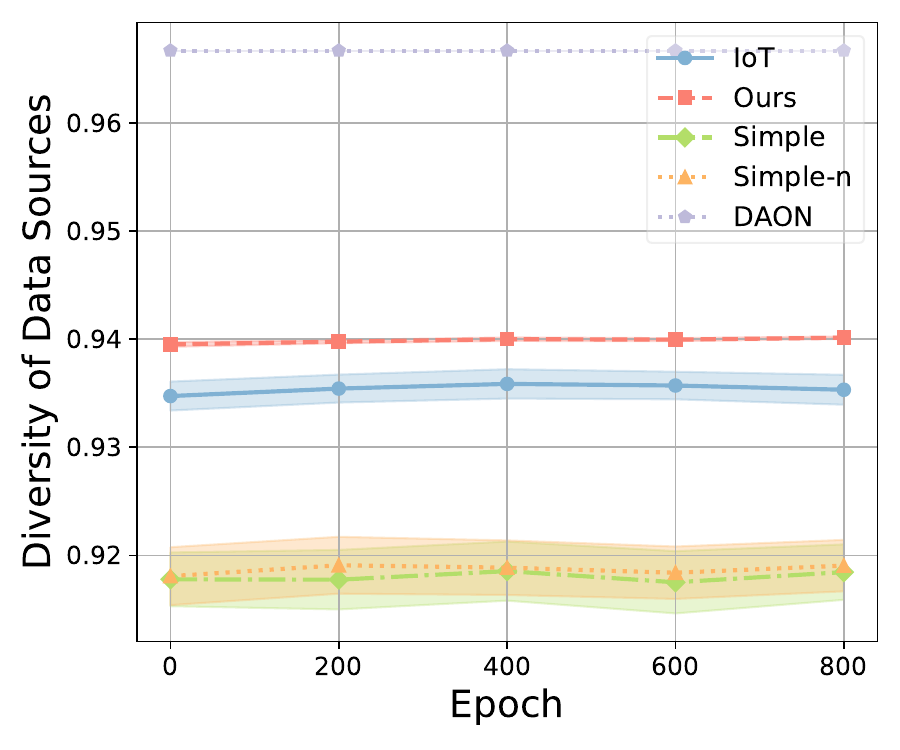}
    \label{subfig:robustness_div}
    }
\caption{Effectiveness in different network environments and network sizes.}
\label{fig:rl}
\end{figure*}

\subsection{Effectiveness Of The Proposed Data Source Selection Strategy}
To analyze the effectiveness of the proposed data source selection strategy, we replace the BLS protocol used by IoT, Simple, etc. schemes with TBLS for comparison. DAON needs to perform aggregation operations at nodes and cannot adapt to TBLS.

Fig. \ref{subfig:tbls_success_rate}-\ref{subfig:tbls_retry_count} illustrates the aggregation success rates and the number of retries for TBLS tasks under the condition that all schemes use the TBLS aggregation protocol and \(K=11\). When the final number of aggregated data sources falls below \(K\), the TBLS protocol will fail and initiate a retry. As the proportion of malicious data sources increases, the aggregation success rate declines while the number of retries escalates. 
Although the IoT approach, which allocates nodes to data sources, does not experience aggregation failures, the on-chain allocation of nodes incurs significant computational and storage overhead. In contrast, our proposed data source selection method shows a clear advantage over Simple-1 approaches and is comparable to Simple-2. Similarly, Fig. \ref{subfig:tbls_time} demonstrates the notable advantage of our proposed scheme in terms of response time.

Fig. \ref{fig:tbls2} illustrates the aggregation success rates and the number of retries for various schemes using TBLS under extreme conditions where \(K=M=20\). First, both the aggregation success rate and the number of retries, as shown in Fig. \ref{subfig:tbls_success_rate_20}, demonstrate results similar to those observed when \(K=11\). Fig. \ref{subfig:tbls_retry_count_20} provides a detailed overview of how the number of retries changes over time. 
As the number of epochs increases, our data source selection strategy gradually learns to select suitable data sources for nodes, thereby avoiding aggregation failures. It indicates that our proposed method not only focuses on improving response times but also enhances the aggregation success rate of TBLS, even in scenarios where complete information about other nodes is not available.

\subsection{Robustness}
To verify the robustness of the proposed scheme, we analyze the effectiveness of the proposed scheme under different network delay distributions and different network sizes.

Fig. \ref{subfig:network_time} and Fig.\ref{subfig:network_div} illustrate the response times and data source diversity of different schemes under network delays following a Gaussian distribution \(N(2.0, 0.4)\), with results resembling those seen in the previous random distribution scenarios. 
Additionally, Fig.\ref{subfig:robustness_time} and Fig.\ref{subfig:robustness_div} demonstrate the performance when the network scale is expanded to 100 nodes and the number of data sources \(M\) increases to 30 while maintaining the thresholds \(t=20\) and \(K=18\). In this environment, {\sysname}'s response time becomes closer to that of IoT, while still maintaining greater diversity than IoT. It indicates the robustness of {\sysname} under varying conditions.

\subsection{Security Analysis}
Based on the security assumptions presented in \$ \ref{security}, we formally analyze how the system achieves the above security goals. Previous research has demonstrated the security of BLS \cite{bacho2022adaptive} and TLS-N \cite{krawczyk2013security}, so we will focus on analyzing how TBLS guarantees data accuracy.

\paragraph{Theorem} If the number of malicious data sources is less than $\frac{M}{2}$ and $K > \frac{M}{2}$ as set by TBLS, {\sysname} will guarantee the accuracy of the final aggregated data.

\paragraph{Proof} Firstly, the nodes exchange data with the data source through the TLS-N protocol, and the proofs generated in this process ensure the integrity of the data and prevent the nodes from tampering with the data. Therefore, even if all nodes $j$ receive data $d'$ from malicious data sources $\mathcal{M}_{\text{malicious}}$, the differences between the data $d'$ provided by these malicious data sources and that provided by the honest data sources lead to the fact that even if $t$ fragments of the same data signatures are collected during the aggregation process $\sigma_{j,d'}$, the diversity requirement $K$ cannot be satisfied due to $\mathcal{M}_{\text{malicious}} < \frac{M}{2} \leq K$, thus preventing successful aggregation. TBLS effectively prevents the influence of malicious data sources on the data aggregation process.

\section{CONCLUSION}
\label{conclusion}
This paper presents a secure and efficient multi-data source oracle solution, {\sysname}, which reduces resource consumption and response times while ensuring data source diversity. To build a low-cost, distributed trust system between nodes and data sources, we design a novel off-chain data aggregation protocol TBLS. Then, under the assumption of rational agents, we model the process of node data source selection as a Bayesian game and apply reinforcement learning to solve it. This approach maximizes node utility while minimizing system resource consumption and response times. Both experimental results and security analysis validate the reliability and effectiveness of the proposed solution.

\section*{Acknowledgments}
The research was supported in part by the National Natural Science Foundation of China (Nos.62166004, U21A20474, 62262003), the Guangxi Science and Technology Major Project (No.AA22068070), the Basic Ability Enhancement Program for Young and Middle-aged Teachers of Guangxi (No.2022KY0057, 2023KY0062), Innovation Project of Guangxi Graduate Education (Nos. XYCBZ2024025), the Key Lab of Education Blockchain and Intelligent Technology, the Center for Applied Mathematics of Guangxi, the Guangxi "Bagui Scholar" Teams for Innovation and Research Project, the Guangxi Talent Highland Project of Big Data Intelligence and Application, the Guangxi Collaborative Center of Multisource Information Integration and Intelligent Processing.

\bibliographystyle{IEEEtran}
\bibliography{IEEEabrv,myref}

\begin{IEEEbiography}[{\includegraphics[width=1in,height=1.25in,clip,keepaspectratio]{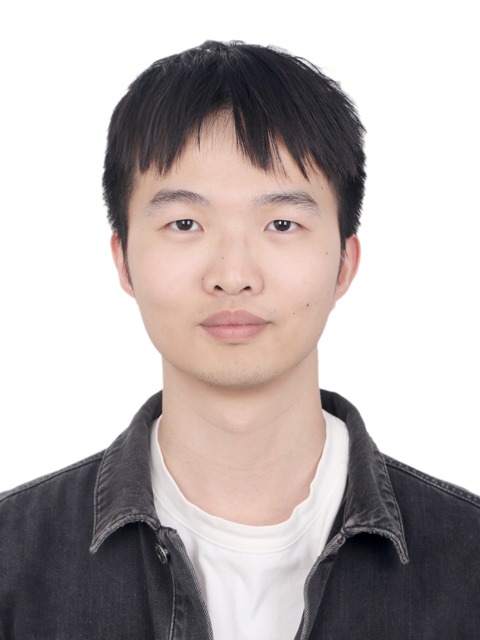}}]{Youquan Xian}
received his master's degree from Guangxi Normal University in 2024. He has published multiple papers in journals and conferences such as IEEE Transactions on Network and Service Management, IEICE Transactions on Communications, IEEE SMC 2024, and WASA 2024. His main research includes blockchain and federated learning.\end{IEEEbiography}

\begin{IEEEbiography}[{\includegraphics[width=1in,height=1.25in,clip,keepaspectratio]{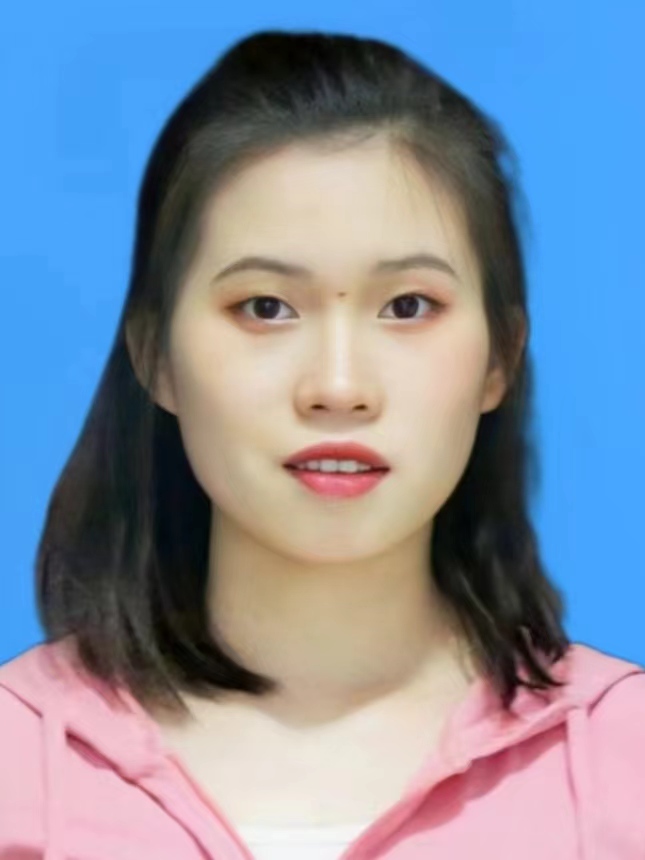}}]{Xueying Zeng} received her bachelor's degree from Guangxi Science and Technology Normal University in 2022. She is currently pursuing her master's degree from Guangxi Normal University. Her research interests are blockchain, crowdsourcing.\end{IEEEbiography}

\begin{IEEEbiography}[{\includegraphics[width=1in,height=1.25in,clip,keepaspectratio]{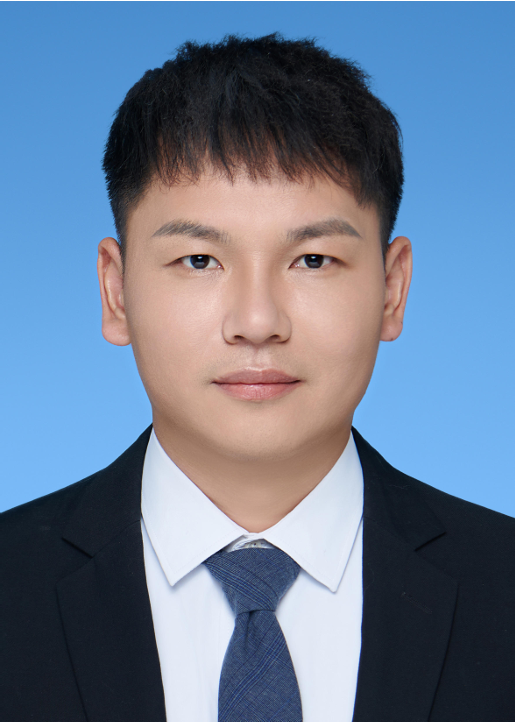}}]{Chunpei Li}
received his Ph.D. from the School of
 Computer Science and Engineering at Guangxi Normal University in 2024. He is currently conducting
 postdoctoral research at the Ministry of Education
 Key Laboratory of Educational Blockchain and Intelligent Technology at Guangxi Normal University.
 His research interests include blockchain, artificial
 intelligence, and information security.\end{IEEEbiography}

\begin{IEEEbiography}[{\includegraphics[width=1in,height=1.25in,clip,keepaspectratio]{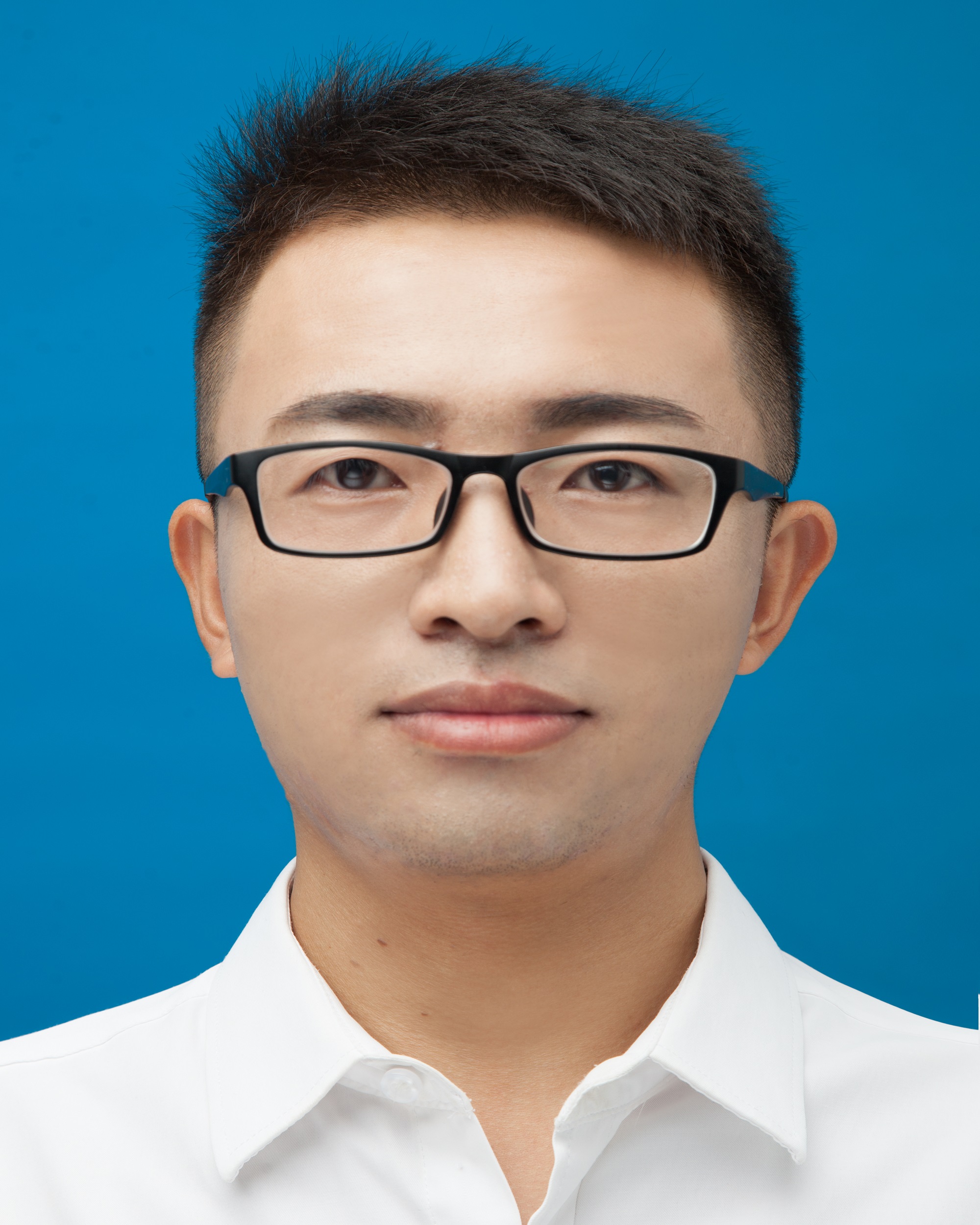}}]{Peng Wang}
received his master's degree from Guilin University of Technology in 2018. He is currently working toward a doctor's degree at Guangxi Normal University. His research interests include blockchain, data fusion, and data security.\end{IEEEbiography}

\begin{IEEEbiography}[{\includegraphics[width=1in,height=1.25in,clip,keepaspectratio]{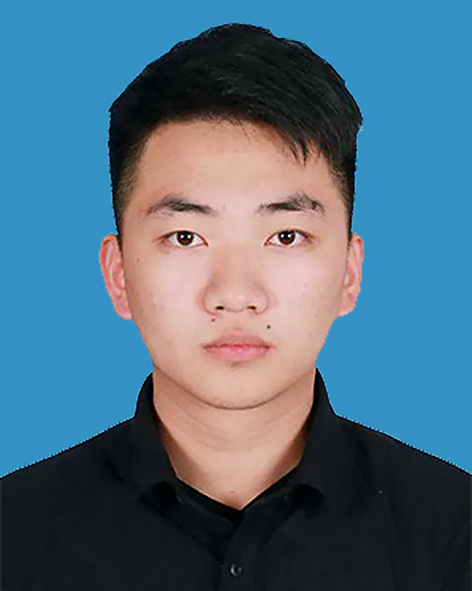}}]{Dongcheng Li}
received his master's degree in sofrware engineering from  Guangxi normal university. He  is currently working at the department of 
Computer Science and Engineering of
Guangxi normal university, 
China. His main research interests include blockchain, data security and recommendation system.\end{IEEEbiography}

\begin{IEEEbiography}[{\includegraphics[width=1in,height=1.25in,clip,keepaspectratio]{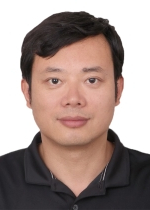}}]{Peng Liu}
received his Ph.D. degree in 2017 from Beihang University, China. He began his academic career as an assistant professor at Guangxi Normal University in 2007 and was promoted to full professor in 2022. His current research interests are focused on federated learning and blockchain.\end{IEEEbiography}

\begin{IEEEbiography}[{\includegraphics[width=1in,height=1.25in,clip,keepaspectratio]{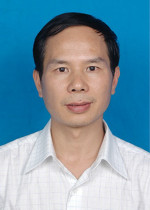}}]{Xianxian Li}
received his Ph.D. degree from the School of Computer Science and  Engineering,  Beihang University, Beijing, China, in 2002. He worked as a professor at Beihang University during 2003-2010. He is currently a professor with  the School of Computer Science and Engineering, Guangxi Normal University, Guilin, China. His research interest includes information security.\end{IEEEbiography}

\vfill

\end{document}